\documentclass[letterpaper,11pt]{article}

\usepackage{ amsmath, epsfig, amssymb,array}

\begin{document}

\baselineskip=18pt


\thispagestyle{empty}
\vspace{20pt}
\font\cmss=cmss10 \font\cmsss=cmss10 at 7pt

\begin{flushright}
\today \\
UMD-PP-09-057,\  MADPH-09-1547,\ NPAC-09-14 \\
\end{flushright}

\hfill
\vspace{20pt}

\begin{center}
{\Large \textbf
{LHC Signals for Coset Electroweak Gauge Bosons \\
in Warped/Composite PGB Higgs Models}
}
\end{center}

\vspace{15pt}

\begin{center}
{\large Kaustubh Agashe$\, ^{a}$, Aleksandr Azatov$\, ^{a}$, Tao
Han$\, ^{b}$, Yingchuan Li$\, ^{b}$, Zong-Guo Si$\, ^{c}$,
Lijun Zhu$\, ^{a}$ } \\
\vspace{15pt}
$^{a}$\textit{Maryland Center for Fundamental Physics,
     Department of Physics,
     University of Maryland,
     College Park, MD 20742, U.S.A.}
\\
$^{b}$\textit{Department of Physics, University of Wisconsin, Madison, WI 53706, U.S.A}
\\
$^{c}$\textit{Department of Physics, Shandong University,
Jinan Shandong 250100, China}

\end{center}

\vspace{20pt}
\begin{center}
\textbf{Abstract}
\end{center}
\vspace{5pt} {\small \noindent The framework of a warped extra
dimension with the Standard Model (SM) fields propagating in it is a
very well-motivated extension of the SM since it can address both
the Planck-weak and flavor hierarchy problems of the SM. Within this
framework, solution to the little hierarchy problem motivates
extending the SM electroweak (EW) $5D$ gauge symmetry in such a way
that its breakdown to the SM delivers the SM Higgs boson. We study
signals at the large hadron collider (LHC) for the extra EW (called
coset) gauge bosons, a fundamental ingredient of this framework. The
coset gauge bosons, due to their unique EW gauge quantum numbers
[doublets of $SU(2)_L$], do not couple at leading order to two SM
particles. We find that, using the associated production of the
charged coset gauge bosons via their coupling to bottom quark and a
(light) Kaluza-Klein excitation of the top quark, the LHC can have a
$3\sigma$ reach of $\sim 2\  (2.6)$ TeV for the coset gauge boson
masses with $\sim 100\ (1000)$ fb$^{ -1 }$ luminosity. Since current
theoretical framework(s) suggest an {\em indirect} lower limit on
coset gauge boson masses of $\stackrel{>}{\sim} 3$ TeV, luminosity
or energy upgrade of LHC is likely to be crucial in observing these
states. }

\vfill\eject
\noindent


\section{Introduction}

The framework of a warped extra dimension with Standard Model (SM) fields propagating in it
\cite{Randall:1999ee, Gherghetta:2000qt}
can address both the Planck-weak and flavor hierarchy problems of the SM:  for a review and further
references, see \cite{review}.
The resolution of the Planck-weak hierarchy relies on the
anti-de Sitter (AdS) geometry leading to {\em exponential} dependence of the effective $4D$ mass scale (including UV cut-off) on the location in the extra dimension.
In particular, this mass scale can be Planckian near one end of the extra dimension
(called Planck brane), where the $4D$ graviton is automatically localized thus
accounting for the strength of gravity. On the other hand,
the natural mass scale can be much smaller, for example,
$O(\hbox{TeV})$, near the other end (called the TeV brane) where the SM Higgs sector
is localized.
Thus, the Higgs mass scale is not sensitive to the Planck scale.
Crucially, such a large hierarchy of mass scales at the two ends of the extra dimension
can be achieved with only a modest  proper size of the extra dimension
in units of the AdS curvature radius.

However, with the 5D electroweak (EW) gauge symmetry being $SU(2)_L\times
SU(2)_R$,\footnote{Here, we include a $SU(2)_R$ factor, which is
motivated by suppressing contributions to the $T$ parameter
\cite{Agashe:2003zs}, as part of the ``SM'' EW gauge symmetry.} and
a Higgs transforming as a bi-doublet (henceforth called ``minimal
Higgs sector''),
%
%
the framework still suffers
from an
%
%
incarnation of the little hierarchy problem.
Namely, the
Higgs mass is still sensitive to the $5D$ cut-off, albeit warped-down
(compared to the fundamental $5D$ scale which is Planckian) at the
TeV brane.
%
%
The problem is that the
mass scale of the Kaluza-Klein (KK) excitations of the SM particles is constrained to be at least a few TeV by various precision
tests (see reference \cite{review}
for a review) and the (warped-down) $5D$ cut-off should be larger than the KK scale
by (roughly) an order-of-magnitude in order for the $5D$ effective field theory description to
be valid.

This naturalness problem
motivates incorporating more structure in (i.e., a non-minimal) \\ Higgs/EW sector.
The idea is to suitably
extend the $5D$ EW gauge symmetry beyond the SM -- the additional $5D$ EW gauge fields are called
coset gauge bosons -- and break it down to the SM by a scalar vev localized near TeV brane
\cite{Contino:2003ve}.
It can be shown that in this process, a
massless (at tree-level) scalar mode (localized near the TeV brane) with SM Higgs quantum
numbers can emerge. Moreover, the quantum corrections to the Higgs mass in this case
has a
reduced sensitivity to the $5D$ cut-off.

In this paper, we begin a study of signals for coset gauge bosons in this framework
at the large hadron collider
(LHC).
We find that the $3 \sigma$ reach of the LHC for the coset gauge
bosons is $\sim 2.6$ TeV with $\sim 1000$ fb$^{-1}$ of integrated luminosity,
under certain well-motivated assumptions which we discuss. However,
we also argue that the (indirect) lower bound on masses of coset
gauge boson masses is expected to be
(at least) $\sim 3$ TeV \cite{Carena:2006bn}
(for review see reference \cite{review}). So, our results provide a strong
motivation for LHC luminosity and/or energy upgrade.

An
outline of our paper is as follows.
We
begin with an overview of
the above framework which we call (in its full generality) ``warped/composite PGB Higgs'' for reasons
which we explain there.
Then, in section \ref{sec:two-site} we present
a discussion of this
%
%
framework using
the convenient ``two-site'' approach \cite{Contino:2006nn} in order to get a {\em general} idea of spectrum and
couplings of coset gauge bosons.
In Section \ref{GHU}, we review specific warped extra dimensional models, namely, minimal ``gauge-Higgs unification'' (GHU) models,
and the mechanism of radiative
%
%
generation of
Higgs potential. In particular, in section \ref{couplings},
we focus on the couplings of
coset gauge boson in the GHU framework,
showing in section \ref{estimate} that the couplings of
coset gauge bosons follow a general pattern which is independent of
the details of this $5D$ model, and then in section \ref{exact}
presenting the {\em exact} formulae for them in the specific
model within this framework by Medina et. al.
\cite{Medina:2007hz}. In Section \ref{numerical}, we show our numerical results
for the particle spectrum and couplings from a scan of parameter
space in the model (in the process
backing-up our {\em estimates} for the couplings
of the coset gauge bosons from section \ref{estimate}), and present sample points for collider study.
Section \ref{signal} focuses on the collider phenomenology, where we study the
production and decay of coset gauge bosons, and the prospect of
their discovery at LHC. We conclude in Section \ref{conclude}.
Technical details of the $5D$ model are relegated to appendices.

\section{Overview}

As discussed in the introduction, we study the warped extra dimensional models
where the SM Higgs arises from the breaking of an extended EW gauge symmetry
down to the SM gauge symmetry near the TeV brane.
A particular limit of this framework is where the scalar vev
involved in this breaking of EW gauge symmetry is much larger than
the AdS curvature
%
%
scale such that the above breaking of
$5D$ EW gauge symmetry is effectively the result of Dirichlet
boundary condition on the TeV brane. The massless scalar mode can
then be thought of as the extra polarization ($A_z$) of the coset
gauge fields. Hence, this model is dubbed ``gauge-Higgs unification
(GHU)'': see, for example, reference \cite{Serone:2009kf} for a
review of and more references for this idea.
Quantum corrections do generate a potential (including a mass term)
for it -- this is
the Hosotani mechanism for symmetry breaking \cite{Hosotani:1983xw}.
However, such effects are saturated at the typical KK scale
rather than at the warped-down $5D$ cut-off
\cite{Contino:2003ve, Agashe:2004rs, Hosotani:2008tx}.

By the AdS/CFT correspondence \cite{Maldacena:1997re}, the general
$5D$ framework mentioned above [i.e., whether vev breaking $SO(5)
\rightarrow SO(4)$ is infinite as in GHU or not] is conjectured to
be a dual description of ($4D$) Georgi-Kaplan (GK) models \cite{GK}. In
GK models,
the SM
Higgs is a composite of purely $4D$ strong dynamics which is also a
pseudo-Goldstone boson (PGB) of a spontaneously broken global
symmetry and hence naturally lighter than the compositeness  scale
(dual to the typical KK scale)
\cite{Contino:2003ve,Arkani-Hamed:2000ds}.
This aspect of the $5D$ models motivates using the terminology
warped/composite PGB Higgs for this {\em general} framework, i.e., including various
$5D$ models [i.e., both the infinite scalar vev for
$SO(5) \rightarrow SO(4)$ breaking, i.e. the GHU models,
and the finite scalar vev] and $4D$ models based on strong dynamics.

Our goal is to study how to distinguish the possibility of such a framework from
the minimal Higgs sector framework by directly producing the extra particles
(i.e., those
arising as a result of the extension of the $5D$ EW gauge symmetry)
at the LHC\footnote{Alternatively, one can probe the extra states indirectly, for example,
via their virtual effects on lower-energy observables or how the properties of the {\em usual}
states are modified in warped/composite PGB Higgs framework
relative to minimal Higgs sector framework. However, such indirect effects
might not be able to provide robust distinction between the two frameworks. The reason is that
the minimal Higgs sector framework has a large number of free parameters
and hence, for some choice of these, can mimic effects of extra particles of warped/composite
PGB Higgs framework.}.
Clearly, the $5D$ fermions -- whose zero-modes are identified with the
SM fermions
-- must also now be in representations of the extended $5D$ EW gauge
symmetry, i.e.,
they are larger than in the case of
minimal Higgs sector, with the extra components not
having zero-modes
(just like the coset gauge bosons).
In particular, in some $5D$ Warped/composite PGB Higgs models,
some of these fermionic KK
states (associated with top/bottom quarks) are lighter than SM gauge KK modes 
\cite{Carena:2006bn, Contino:2006qr, Medina:2007hz}
(and hence, as discussed below, lighter than
the coset gauge boson), whereas
KK fermions have same mass as gauge KK modes in minimal Higgs sector framework.
Hence these fermionic KK modes  might be easier to detect at the LHC \cite{Carena:2007tn}
than the SM (or coset) gauge KK modes
and their discovery would be suggestive of warped/composite PGB Higgs models
rather than the models with minimal Higgs sector.
%
%
%
However, in the models constructed so far, most of these light
fermionic states have the same quantum numbers under SM EW symmetry
as those of SM fermions\footnote{The exception is a 5/3-charged light
KK fermion, but its existence might have more to do with the need for
$Zbb$ protection rather than PGB Higgs. } so that they could be
mistaken for similar states in {\em other} extensions of the
SM\footnote{KK fermionic states in the minimal Higgs sector
framework also have the same quantum numbers, although these
states are expected to be as heavy as SM gauge KK modes. So,
there is less possibility of confusion between minimal Higgs sector
models and warped/composite PGB Higgs framework based on these
states.} . Thus, it is crucial to consider {\em additional} signals
for the warped/composite PGB Higgs framework.

%

Such a test can be provided by detection of
\begin{itemize}

\item
the coset gauge bosons which, being doublets of $SU(2)_L$, have novel  (i.e., non-adjoint)
representations under the SM EW gauge symmetry

\end{itemize}
%
%
%
(such quantum numbers for gauge bosons are obviously absent in the minimal
Higgs sector framework). Thus these coset gauge bosons can result in distinctive LHC signals as compared to
EW gauge KK modes in minimal Higgs sector models.
Similarly,
we discuss how coset gauge bosons can also
be differentiated from new gauge bosons in {\em other} extensions of the SM.
This feature of coset gauge bosons motivates our study in this paper
of signals from their direct production at the LHC.\footnote{Very
recently, in reference \cite{Chizhov:2009fc}, a different signal
(than what we study) for coset gauge bosons was suggested based
(again) on the distinctive quantum numbers, but it was not studied
in the context of a complete framework, for example,
one that explains the flavor
hierarchy.}
%
%

Our study suggests that
\begin{itemize}

\item
the LHC $3 \; \sigma$ reach for (charged) coset gauge bosons masses is
$\sim 2 (2.6)$ TeV with $\sim 100 (1000)$ fb$^{-1}$ luminosity, using
their associated production with (light) KK top and decay into KK top
and bottom quarks.

\end{itemize}
For this analysis, we use values of
couplings
which are motivated by the ($5D$) {\em minimal} (i.e., with no brane-localized kinetic terms
for bulk fields and with AdS$_5$ metric: see later) GHU model.
A note on the allowed mass scale is in order here.
In the
minimal GHU model, it turns out that the coset gauge boson
mass is $\approx 5/3$ larger than SM gauge KK modes \cite{Agashe:2004rs}.
And, the lower bound on the latter gauge boson masses is $\sim 3$
TeV from EW precision tests \cite{Carena:2006bn} (for a review see
reference \cite{review}), assuming
custodial symmetries are implemented \cite{Agashe:2003zs,
Agashe:2006at} (and, depending on details of flavor structure, the
bound can be somewhat stronger from flavor violation
\cite{Huber:2003tu, Csaki:2008zd}\footnote{See references
\cite{Agashe:2009tu} and \cite{Gedalia:2009ws} for ``latest''
constraints from
lepton and quark flavor violation, respectively, i.e.,including
variations of the minimal framework.} although these constraints can
be ameliorated by addition of $5D$ flavor symmetries
\cite{Fitzpatrick:2007sa}).
Thus, the coset gauge boson mass is constrained to be at least $5$
TeV which is well beyond reach of even $1000$ fb$^{-1}$ luminosity
at the LHC.\footnote{The bound on mass scale of coset gauge bosons
from precision tests involving exchange of coset gauge bosons
themselves is rather weak since there is no coupling of single coset
gauge boson to purely SM particles at leading order (simply due to
quantum numbers) so that coset gauge boson exchange at {\em
tree}-level does not contribute to purely SM operators (and hence
precision tests). The flip side of this feature is that resonant
production of coset gauge bosons is suppressed, which is in part
responsible for the poor LHC reach.}
%
%
%
This situation then
motivates upgrade of the energy of the LHC or building another higher-energy collider.

However,
in
{\em non}-minimal $5D$ models -- for example, with brane-localized
kinetic terms for bulk fields \cite{Davoudiasl:2002ua}
or with the metric near the TeV brane being modified from pure AdS \cite{soft} {\em within} the
GHU models or with the scalar vev giving masses to coset gauge bosons
being finite (instead of infinite as in GHU models),
the indirect bound on
coset mass scale might be relaxed because the
ratio of coset to SM gauge KK masses is closer to 1.
In fact, inspired by deconstruction/latticization and the
AdS/CFT correspondence,
a purely $4D$, two-site approach \cite{Contino:2006nn} --
keeping only SM and 1st KK excitations --
has been proposed in
order to efficiently/economically capture the phenomenology of
similar variations of $5D$ models
with a {\em minimal} Higgs sector.
Such a two-site approach
can be extended to PGB Higgs
models as well \cite{private}.

%
%
Using a two-site approach for the general warped/composite PGB Higgs, we argue
that
\begin{itemize}

\item

coset gauge bosons are expected to be at most be as light as (i.e., cannot be
lighter than) SM gauge KK (or composite) modes.

\end{itemize}
Moreover, using the same approach, it can be shown
that the bound on SM gauge KK
(or composite) modes is unlikely to be reduced
below $\sim 3$ TeV even in the non-minimal models, i.e.,
in the general framework\footnote{It has been claimed that
in soft-wall models, this bound can be lower than
$3$ TeV. However, such models have not been developed fully as yet.}
Thus,
coset gauge bosons are expected to have mass $\stackrel{>}{\sim} 3$ TeV
in general.
We argue based on the two-site approach
%
%
%
that
couplings of coset gauge bosons in the general framework
will still be similar to those in minimal $5D$ GHU models which we used for the study of LHC signals.
This feature
implies that the LHC reach for coset gauge bosons that we find based on couplings
in minimal $5D$ GHU model is expected to apply in general to the framework of
warped/composite PGB Higgs. Thus, even optimistically, i.e.,
assuming that in some
models within this framework the coset gauge bosons can be as light as SM gauge KK modes
{\em and} using the $1000$ fb$^{-1}$ luminosity,
we see that the LHC can barely be sensitive to
the lower (indirect) limit of $\sim 3$ TeV on coset gauge boson masses.


\section{Model-independent Analysis Using Two-site Approach}\label{sec:two-site}

In this section, we
provide a {\em rough}
description of masses and couplings of the coset gauge bosons
of the {\em general} warped/composite PGB Higgs framework, i.e.,
the analysis presented here is applicable to both $5D$ and $4D$
models in this framework.
The detailed
description of a {\em specific} (5D) model, namely minimal GHU, will be given in the next
section.

Here we use the two-site model \cite{Contino:2006nn} which is a
convenient parametrization for this framework. It can be shown that
this effective 4D description is the deconstructed version of warped
extra dimension models with SM fields propagating in the bulk,
including the zero and only the 1st KK modes. In the original setup
presented in reference \cite{Contino:2006nn}, Higgs was {\em not} a
PGB. So first we will briefly review this model (for more details,
the reader is referred to this paper), and then we will show what
changes we have to make to account for the PGB origin of the Higgs.

The original two-site model consists of two sectors: ``elementary" and
``composite" (this nomenclature is inspired by the AdS/CFT correspondence).
The elementary sector is a copy of all SM states
except for the Higgs field. The composite sector
consists of
massive gauge bosons, massive vector-like fermions and the Higgs
field. The composite sector states live in complete
representation of the {\em global} symmetry $SU(3)_c\times SU(2)_L\times
SU(2)_R\times U(1)_X$, where the additional custodial $SU(2)_R$ is
introduced to suppress new physics contribution for $T$ parameter.
The massive gauge bosons live in adjoint representation while part
of massive fermions live in the same representation as that of SM
fermion.

These two sectors mix with each other, leading to massless fermion
and gauge boson eigenstates which correspond to SM fermions
($\psi_{L,R}$) and gauge bosons ($A_\mu$) before Electro-Weak
Symmetry Breaking (EWSB). The heavy eigenstates are denoted by
$\rho_\mu$ for gauge bosons and $\chi_{L,R}$ for fermions.\footnote{The SM states and heavy
eigenstates further mix with each other after EWSB,
but this effect is not relevant here.} 
The SM states (except for the Higgs) are mixtures of elementary and
composite states:
\begin{eqnarray}\label{Eq. mixing}
|\text{SM}\rangle = \cos\theta ~| \text{elementary}\rangle +
\sin\theta ~|\text{composite}\rangle ~,
\end{eqnarray}
where all SM states (except for the top) are mostly made of the
elementary sector ones (i.e., $\sin\theta \ll 1$), while the heavy
states are mostly the composite sector ones and finally the SM Higgs
is fully a composite sector state. The composite sector states are
assumed to have strong couplings to each other, in order to match
the 5D description (or equivalently, inspired by AdS/CFT
correspondence). We use $g_*$ and $Y_*$ to denote the composite
gauge and Yukawa couplings (and will take them to be roughly of
order a few). In the flavor anarchy models, $Y_*$ for different
flavors are of the same order and have no structure, which we assume
for the following discussion. Clearly the SM states couple to heavy
states through the mixing (Eq.~(\ref{Eq. mixing})). For example the
Yukawa couplings between SM fermions and Higgs is given
schematically by
\begin{eqnarray}
Y_{SM} \sim \sin\theta_{\psi_L} Y_* \sin\theta_{\psi_R} ~.
\end{eqnarray}
The fermionic mixing angles $\theta_{\psi_{L,R}}$ are assumed to be
hierarchical, which explains the SM fermion mass hierarchy. In
warped extra dimension picture, $\sin\theta_\psi$ is related to the
fermion zero mode wavefunction evaluated at the TeV brane (see
$f(c)$ in Eq.~(\ref{Eq. fc}), with an exponential dependence on 5D
mass parameter, $c$), thus the hierarchical mixing angles
$\sin\theta_\psi$ can be naturally generated in the warped extra
dimension picture.

\begin{figure}
\includegraphics[scale = 0.6]{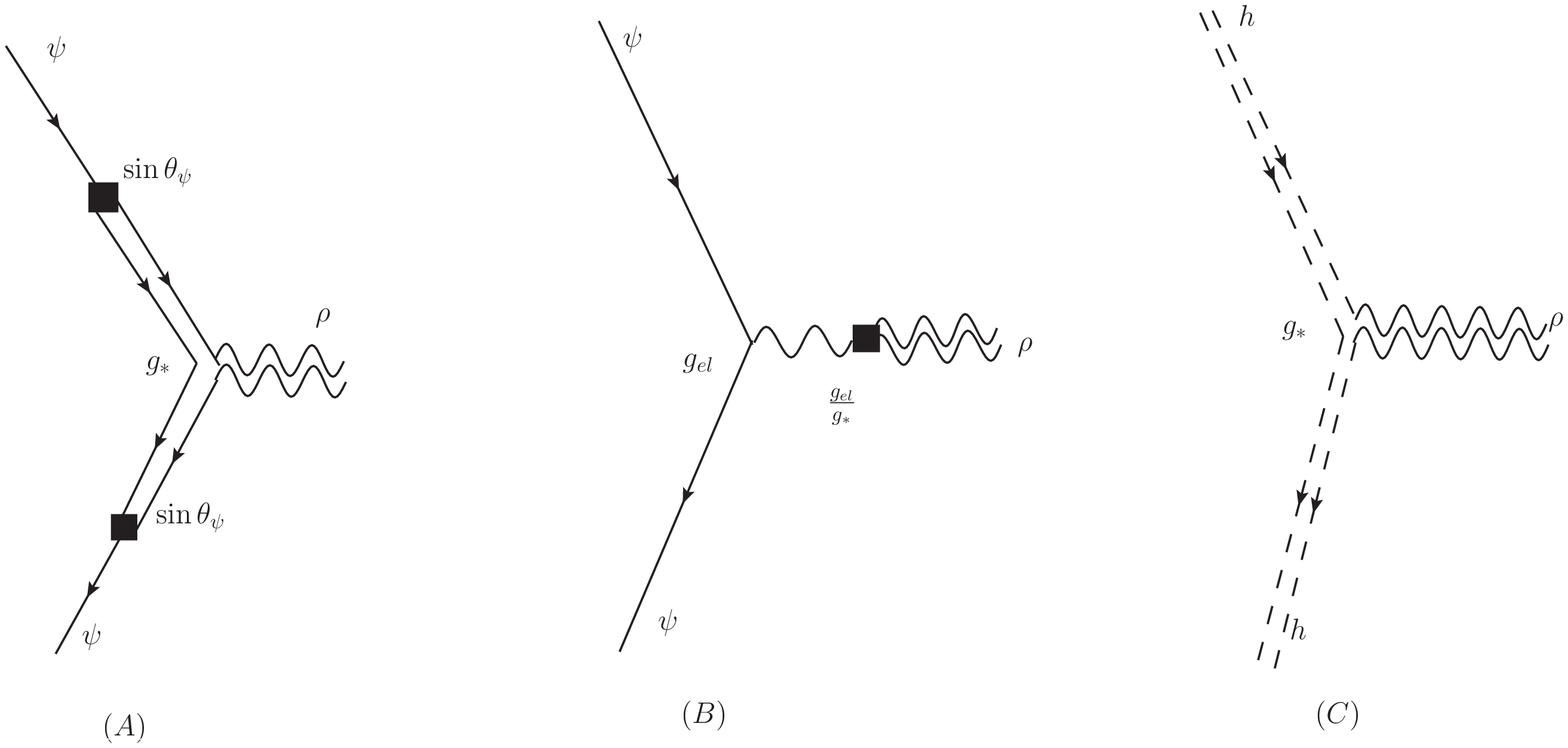}
\caption{Couplings of heavy gauge bosons with SM states. Fig. (A),
(B) give the couplings between heavy gauge bosons and SM fermions
coming from fermionic and gauge boson mixings, respectively. Fig. (C)
gives the coupling between heavy gauge bosons and Higgs field,
which after EWSB give rise to the coupling to physical Higgs and
longitudinal $W/Z$. }\label{Fig. 2siteffg}
\end{figure}

The mixing angles for gauge bosons are given by $\sin\theta_G =
\frac{g_{el}}{g_*}$, where $g_{el}$ is the elementary gauge
coupling, while the SM gauge coupling is given by $g_{ SM } = g_{ el
} g_* / \sqrt{ g_{ el }^2 + g_*^2 }$.
We will choose $g_*\sim$ a
few such that $g_* \gg g_{ SM }$ and thus $g_{ el }$ can be approximated  by
the SM gauge couplings.
Specifically, in order to match 5D theories $\sin\theta_G$ should  be
$\sim 1 / \sqrt{ \hbox{logarithm of UV-IR hierarchy} }$,
i.e., $\sim 1/6$ for the case of Planck-weak hierarchy.
Here we
review the couplings of heavy gauge bosons $\rho_\mu$ to SM states,
which we use to compare with  the couplings of  coset gauge bosons
later. The SM fermion coupling to heavy gauge bosons are generated
both through fermionic and gauge boson mixings. This is illustrated
using insertion approximation in Fig.~\ref{Fig. 2siteffg} (A)(B) .
This gives the coupling
\begin{eqnarray}\label{Eq. rpsipsi coupling}
g_{\rho\psi\psi} \approx \sin\theta_{\psi} g_*
\sin\theta_{\psi}+\frac{g_{el}^2}{g_*} ~.
\end{eqnarray}
Note that there is a flavor dependent contribution (first term in
the above equation) that comes from elementary composite mixing of
fermions, which is suppressed by the fermionic mixing angles
$\sin\theta_{\psi_{L,R}}$, and there is a flavor universal
contribution (second term in the above equation) that comes from
elementary/composite mixing of gauge bosons, which is suppressed by
$\frac{g_{el}}{g_*}$ relative to SM gauge couplings. For light
fermions, the flavor universal term dominates. Moreover, this coupling
is only mildly (i.e., $\sim 1/6$) suppressed relative to the SM one. The heavy gauge
bosons couple strongly to Higgs field since they are both mostly
composite states (See Fig.~\ref{Fig. 2siteffg}(C)):
\begin{eqnarray}\label{Eq. rhh coupling}
g_{\rho h h} \approx g_* ~.
\end{eqnarray}
Using Goldstone boson equivalence theorem, we can see that after
EWSB, the heavy gauge bosons acquire strong coupling $\approx g_*$
with physical Higgs boson and {\em longitudinal} component of $W/Z$.

We now turn to the two-site description of warped/composite {\em
PGB} Higgs. First let us ignore $SU(3)_c\times U(1)_X$ part of the
composite sector global symmetry because it is irrelevant for the
Higgs part of the model. We want composite sector of the model to
have a global symmetry $H$ which includes $SU(2)_L\times SU(2)_R$
[latter group is isomorphic to SO(4)]. At the same time, Higgs should be
a PGB. One can see that in order to achieve this setup, the composite
sector should have  larger global symmetry $G$, which later should
be spontaneously broken down to its subgroup $H$, and  Higgs is PGB
of this symmetry breaking pattern $G\rightarrow H$ in the composite
sector.  The simplest example which we will study in this paper
corresponds to the $G = SO(5)$ and $H = SO(4)$, i.e., the {\em full}
global symmetry of the composite sector is extended from $SU(3)_c
\times SU(2)_L \times SU(2)_R \times U(1)_X$ of the original model
to $SU(3)_c \times SO(5) \times U(1)_X$. One can see that due to the
larger symmetry $G$ of the composite sector there will be additional
heavy gauge bosons which belong to the group $G/H$ (i.e., the
coset), and they correspond to the coset gauge bosons of the general
warped/composite PGB Higgs framework.

We can learn some important properties of the coset gauge bosons
based on this simple setup. First, we argue that coset gauge bosons
are generally heavier than the usual composite gauge bosons (i.e.
the gauge bosons of the gauge group $H$). The argument is the
following. Before the symmetry breaking $G\rightarrow H$, the gauge
bosons of $H \,(\rho^\mu)$ and $G/H \,(\rho^{\mu}_c)$ of the
composite sector should have the same mass (due to the global
symmetry, $G$). After the symmetry breaking, the masses of the gauge
bosons of $H$ remain the same, while the coset gauge bosons in $G/H$
get extra mass contribution coming from the breaking. For example,
for the case that we consider, i.e., with $G = SO(5)$ and $H =
SO(4)$, the breaking $G\rightarrow H$ can be achieved by the vev of
a scalar $\phi$ transforming in fundamental representation of
$SO(5)$. We can parameterize $\phi$ by
\begin{eqnarray}
\phi = e^{-i T^a_c h^a}\left( \begin{array}{c} 0\\ 0 \\ 0 \\ 0 \\
f_\phi + \eta
\end{array} \right)~,
\end{eqnarray}
where $f_\phi$ is the magnitude of $\phi$ vev, $T_c^a$ are the
generators of $G/H$, $h^a$ are the pseudo-Goldstone bosons which are
also the Higgs, $\eta$ is a massive scalar excitation. The covariant
derivative of $\phi$ gives rise to extra contribution to the masses
of coset gauge bosons
\begin{eqnarray}\label{Eq: fphi}
(D_\mu \phi)^\dagger(D^\mu \phi) \supset g_*^2 f_\phi^2 \rho_{c,\mu}
\rho_c^\mu ~.
\end{eqnarray}
This extra contribution is always positive, thus $\rho_c^\mu$ are
generally heavier than $\rho^\mu$\footnote{assuming only the minimal couplings of
$\phi$ to coset gauge bosons as above.}. We conclude that
\begin{itemize}

\item
there is an
indirect bound of $\gtrsim 3$ TeV for the coset gauge boson masses,
which comes from the bound (from precision tests) of $\sim 3$ TeV on
the ordinary composite gauge bosons (as mentioned earlier).

\end{itemize}
On the
other hand, since it is the coset gauge bosons that cancel the
quadratic divergence in Higgs mass from $W/Z$ loops, it is clear
that naturalness favors the coset gauge bosons to not be heavier
than several TeV.
We can also study the structure of coset gauge boson {\em couplings}
based on the two-site language. Note that the discussion here is
independent of the scale $f_\phi$ (see Eq.~(\ref{Eq: fphi})) that
controls the masses of coset gauge bosons (relative to the other
gauge bosons)\footnote{$f_{ \phi}$ is also (roughly) related to the
size of the scalar vev breaking $SO(5) \rightarrow SO(4)$ in the 5D
model.}. We will see that the quantum numbers of coset gauge bosons
give important restrictions on their couplings. First, we study
their couplings with two SM gauge bosons. For this purpose, we
consider the SM gauge bosons before EWSB: $W^a_\mu$ transform in
adjoint representation of $SU(2)_L$ and $B_\mu$ transform as a
singlet. The SM quantum numbers of coset gauge bosons are the same
as that of Higgs, i.e., they are $SU(2)_L$ doublet. Just based on
quantum numbers, we can see that
\begin{itemize}

\item
there is no coupling between one coset gauge boson and two SM gauge
bosons or two Higgs bosons at lowest order (i.e., without EWSB),

\end{itemize}
which is independent of the
elementary/composite nature of SM/coset gauge bosons. This is to be
contrasted with Eq.~(\ref{Eq. rhh coupling}) and Fig.~\ref{Fig.
2siteffg}(C), where we see that the usual heavy gauge bosons have
large couplings to Higgs bosons and longitudinal $W/Z$.

We turn to the couplings between coset gauge bosons and SM fermions.
We denote the SM fermions by $q_L$ ($SU(2)_L$ doublet) and $u_R$
($SU(2)_L$ singlet) respectively, where $L,R$ subscripts stand for
the 4D chirality.\footnote{Couplings of coset gauge bosons to
right-handed down-type quarks and leptons can be similarly studied,
but these states are not relevant
%
%
here
%
%
since the associated elementary-composite mixings (even for bottom quark
and $\tau$, i.e., the heaviest fermions) are small
and, as we will discuss later, these sectors also do not
result in light KK states.}
Based on quantum numbers, we cannot write down dimension 4 coupling
between SM fermions and coset gauge bosons. The only allowed
dimension 4 couplings are
\begin{eqnarray}\label{eq. coset fermion two site}
g_{ q U } \bar{q}_L \gamma_\mu \rho_{c}^\mu U_L + g_{ u Q } \bar{u}_R \gamma_\mu
\rho_c^\mu Q_R + h.c. ~, \label{qn1}
\end{eqnarray}
where $Q_R, U_L$ are heavy (purely composite) fermionic
states transforming under
$SU(2)_L$ as doublet and singlet, respectively, i.e.,
opposite chirality to the SM fermions.
Recall that the composite sector fermions are in vector-like
and complete
representations of $SO(5)$ (in particular, the
SM gauge group), while the elementary sector fermions are only in
complete, chiral representation of SM gauge group.

There could be higher dimensional
operators that couple coset gauge bosons with just SM fermions.
These couplings can be schematically written as:
\begin{eqnarray}\label{eq coset gauge bosons SM fermions}
\frac{\tilde{g}_q}{\Lambda} \bar{q}_L \gamma_\mu \rho_{c}^\mu h q_L
+
\frac{\tilde{g}_u}{\Lambda} \bar{u}_R \gamma_\mu \rho_c^\mu h u_R ~,
\label{qn2}
\end{eqnarray}
where $\Lambda$ is some mass scale which depends on the specific
%
%
model. There could also be magnetic dipole moment type operators
involving just SM fermions and coset gauge bosons:
\begin{eqnarray}\label{eq: dipole}
 \frac{g_{\text{dipole}}}{\Lambda} \bar{q}_L \sigma^{\mu\nu}
 D_{[\mu}\rho^c_{\nu]} u_R + \text{h.c.},
\end{eqnarray}
where $D_\mu$ is the covariant derivative operator with respect to
$SU(2)_L\times U(1)_Y$.

\begin{figure}
\hspace{2 cm}
\includegraphics[scale = 0.45]{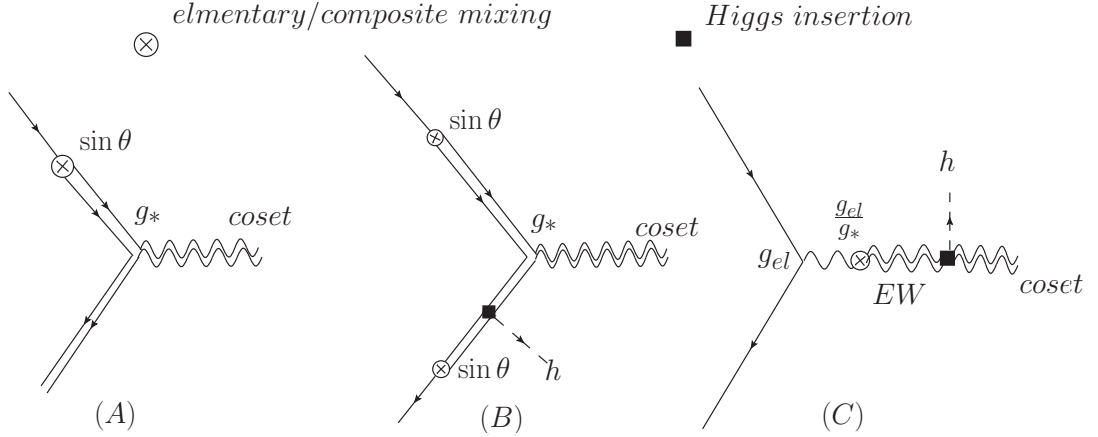}
\caption{Couplings between coset gauge bosons and fermions using
insertion approximation. Fig. (A) shows the couplings between coset
gauge boson, SM fermion and composite fermion.  Estimates of these
couplings are given in Eq. (\ref{g estimate}). Fig. (B) shows the
couplings between coset gauge boson and two SM fermions coming from
elementary/composite mixing of fermions. A Higgs insertion is needed
since otherwise the composite fermion cannot mix with elementary
fermion due to quantum numbers, namely, this composite fermion has
opposite-to-SM chirality. Estimates of these couplings are given in
Eq. (\ref{g sm estimate1}). Fig. (C) shows the couplings between
coset gauge boson and two SM fermions coming from the mixing of
elementary and
composite gauge bosons of the
SM-type (denoted by ``EW'')
followed by their mixing with coset gauge bosons induced by the
Higgs vev. Estimates of these couplings are given in  Eq. (\ref{g sm
estimate2}).}\label{Fig two-site-coupling}
\end{figure}

So far we have been analyzing the couplings of the coset gauge
bosons  based only on their quantum numbers without implementing any
specific property of the model. Let us now first estimate the size
of the couplings ${g}_{q U,u Q }$ in Eq.~(\ref{qn1}) based on our
two-site description of the general warped/composite PGB Higgs
framework, i.e., utilizing the elementary or composite sector nature
of the various particles. Since the SM fermions are mostly
elementary, the above couplings must arise due to
elementary/composite fermionic mixing. In the insertion
approximation  the couplings in Eq.~(\ref{qn1}) will be generated
(dominantly)
from the diagram shown in Fig.\ref{Fig two-site-coupling}(A) and
thus can be estimated to be:
\begin{eqnarray}\label{g estimate}
g_{ q U } \sim g_* \sin\theta_{q_L}, \quad g_{ u Q } \sim g_* \sin\theta_{u_R},
\end{eqnarray}
For the third generation quarks (especially top quark), it is
possible that $\sin\theta_{t_{L,R}}, \sin\theta_{b_L} \sim O(1)$.
Therefore,
\begin{itemize}

\item
coset gauge boson should couple strongly with $t_{L,R},
b_L$ and composite fermions.

\end{itemize}

Turning now to the couplings in Eq.~(\ref{qn2}), they
%
%
can be
generated via elementary-composite fermion mixing,
as shown in Fig.\ref{Fig
two-site-coupling}(B) in insertion approximation. Thus the mass scale $\Lambda$ of Eq.
(\ref{qn2}) in the two-site description of PGB Higgs becomes equal
to the mass of composite sector fermions (denoted by $M_*$) and the
order of the coupling factor $\tilde{g}$ can be estimated to be $g_*
\sin \theta_{ \psi }^2$.
%
%
Once Higgs gets vev, the couplings between coset gauge bosons and SM
fermions can then be generated:
\begin{eqnarray}\label{g sm estimate1}
g_{ q q }^{ f.d. }\sim \frac{v}{ M_*
%
%
} g^2_* (\sin\theta_{q_L})^2, \quad
g_{ u u }^{ f.d. }\sim \frac{v}{ M_*
%
%
} g^2_* (\sin\theta_{u_R})^2,
%
%
%
%
%
\end{eqnarray}
where ``f.d.'' denotes flavor-dependent couplings.

Another contribution to these couplings comes from elementary-composite
{\em SM}-type gauge boson mixing -- recall that there is
no elementary coset gauge boson, followed by composite SM-coset mixing via
Higgs vev, as shown in Fig. \ref{Fig two-site-coupling}(C) with
\begin{eqnarray}\label{g sm estimate2}
g^{ f.i.}_{ q q, \; u u } & \sim & \frac{ g_{ el }^2 }{ g_* } \frac{ g_* v }{ M_* }
\end{eqnarray}
Clearly, these couplings are flavor-independent (hence
denoted by ``f.i.'') and dominate the ones in
Eq. (\ref{g sm estimate1}) for light SM fermions, whereas those in Eq. (\ref{g sm estimate1})
dominate for third generation
quarks.
The magnetic dipole moment type operator (Eq.~(\ref{eq: dipole})) is
recently discussed in \cite{Chizhov:2009fc}. In our framework, this
operator is only generated through loop processes, and it is further
suppressed by the fermion mixing angle $(\sin\theta_\psi)^2$.
Therefore, it is not phenomenologically important here, even for
top/bottom quarks.

We can now study the phenomenological implications of these couplings of coset gauge bosons.
As studied in \cite{Agashe:2007ki}, the dominant {\em production}
channel for the ``usual''
(i.e., transforming as adjoint of SM gauge
group) heavy gauge bosons is through the
(flavor-universal) coupling between light quarks and heavy gauge
bosons (see second term of Eq.~(\ref{Eq. rpsipsi coupling}) and
Fig.~\ref{Fig. 2siteffg}(B)).
However, as argued above,
the coupling of coset gauge bosons
to two light quarks
is suppressed by $\sim g_* v / M_*$ compared to similar couplings of usual heavy gauge bosons.
Since for realistic models one usually finds $g_*v / M_* \lesssim 0.4$,
the
resonant production of coset gauge bosons via light quarks
is expected to
be suppressed by at least an order of magnitude compared to that of usual heavy gauge bosons (for the same mass).
%

On the other hand, the dominant discovery
channel for usual heavy gauge boson is via {\em decay} into two Higgs or
two (longitudinal) W/Z gauge bosons or into two third generation quarks due to the
composite sector nature of all these particles. However, the main decay channels for coset gauge
bosons
are
into one third
generation SM quark and one {\em heavy} quark based on
above analysis; of course, for this decay to be kinematically allowed, the heavy
quark must be lighter than coset gauge bosons -- we find that such a scenario does
indeed occur in part of the parameter space\footnote{The coupling of coset gauge
bosons to {\em two} heavy (mostly composite) fermions is also large, but
we find that (typically) such a decay is not kinematically possible and hence this
coupling is not relevant for our analysis.}.
Again, even if $\sin\theta_{t_{L,R},b_L} \sim O(1)$, the decay into
two third generation SM quarks is suppressed compared to the decay
into one third generation quark and one heavy quark due to suppression of
the former coupling by $\sim g_* v / M_*$ relative to the latter.
And, couplings of coset gauge bosons to light quarks are even smaller.
%
%

We can see that the phenomenology of coset
gauge bosons is very distinct from that of usual
heavy gauge bosons so that the two types of gauge bosons can be
distinguished based on their signals at the LHC. Note that the conclusion here is general in the
sense that it is the result of the quantum numbers of coset gauge
bosons and their (purely) composite sector
nature. This renders our collider study
to be robust and not dependent on specific models.

\subsection{ ``Pollution'' from the usual heavy gauge bosons in the
signal from the resonant production of coset gauge bosons}

Finally, we would like to emphasize that a study of channels other
than resonant production (via light quarks) for coset gauge bosons
is motivated for the following reasons.
The point is that for resonant production of coset gauge bosons, even though the coset gauge bosons
have distinctive decays (as discussed above), it turns out in the end that there is
a larger contribution from resonant production of the usual heavy gauge bosons
(i.e., composite sector $W/Z$'s) to the same final states.
Given that coset gauge boson is a doublet of $SU(2)_L$, it is clear that the dominant fermionic decay (i.e.,
not requiring EWSB) of coset gauge bosons is to a doublet and singlet
(whether SM or heavy)
-- we will focus here on final state with SM top/bottom and composite (heavy, i.e.,
non-SM) fermion.
Whereas,
the usual heavy gauge bosons are triplets/singlets and so
cannot decay without EWSB into this final fermionic state, but instead decay into two doublets or two singlets.
However, after EWSB, the usual heavy gauge bosons can decay into the same final state
as the coset gauge boson.
One possibility is EWSB mixing on {\em gauge boson} line, i.e.,
the usual heavy gauge bosons do have an admixture of coset gauge bosons
[again, resulting from the Higgs vev as
shown in Fig. \ref{Fig two-site-coupling}(C)].
Hence, via this coset gauge boson component, the usual heavy gauge bosons
will decay into the same final state
as that of the coset.
Of course, this effect does not really constitute a ``pollution'' since it does
require presence of the coset gauge boson, i.e., in this case, the top/bottom
and composite fermion final state can
still be taken as ``evidence'' for coset gauge boson.

However, another possibility is EWSB mixing on {\em fermion} line: the composite sector $W/Z$'s
decay into two doublets or two singlet fermions, followed by doublet mixing with a singlet (or vice versa)
via EWSB, i.e., the fermionic mass eigenstates are also admixtures of doublet and singlet.
The crucial point is that this decay of usual heavy gauge boson to the same final
state as that of coset gauge boson
can occur even in the {\em absence} of the coset gauge boson and hence is a genuine pollution.
Of course, such decays
of usual heavy gauge bosons will be
suppressed by this EWSB mixing, i.e., factors of
$g_* v / M_*$ (or $Y_* v / M_*$), compared to other final states
such as $WW/WZ$ and $t \bar{t}/t \bar{b}$ to which the usual heavy gauge bosons
couple strongly.
However, it is clear that the above
suppression in the decay of usual heavy gauge
bosons to the same  final state
as for coset gauge boson
simply serves to compensate (in the total amplitude)
the larger coupling (as mentioned above) of the usual heavy gauge bosons
to the initial state light quarks.
Moreover,
given that
the coset gauge bosons are heavier than usual gauge bosons
(they cannot be lighter as suggested by the two-site description),
the PDF's will then result in the
contribution to the top/bottom and composite fermion final state from
the production/decay of usual heavy
gauge bosons actually {\em dominating} that from coset gauge bosons.

Thus, in this case, the pollution from
usual heavy gauge bosons might make it
difficult to extract a signal for the coset gauge bosons from their
resonant production via light quark annihilation.
Of course, we could undertake the difficult task of
reconstruction of the invariant mass of the final state of top/bottom and composite fermion in order to separate
the two contributions (again, coset gauge bosons are generically heavier than the usual heavy gauge bosons).
Thus, a very
careful study (i.e., including
production of usual heavy gauge bosons), which is
beyond the scope of this paper, would be required
to ascertain whether resonant production via light quarks is actually
a useful channel.
Therefore,
in section \ref{signal},
we will pursue another channel (namely associated production of
$W_C$) which has comparable cross-section to resonant production
via light quarks and furthermore has no significant pollution
from production of the usual heavy gauge bosons.

%


\section{Gauge-Higgs Unification in Warped Extra Dimension}
\label{GHU}

\subsection{Gauge Bosons and Higgs Fields}

Having discussed general two-site description of the general
warped/composite PGB Higgs framework, we now turn to a specific 5D
model. In this section, we review models of (minimal) GHU in a
warped extra dimension: for more details, see reference 
\cite{Medina:2007hz} whose notation we will mostly follow 
here (see also references
\cite{Contino:2006qr, Csaki:2008zd} for similar analyses).
The spacetime metric is given by
\cite{Randall:1999ee}
\begin{equation}
ds^2 = \frac{1}{(kz)^2} \left\{ \eta_{\mu\nu} dx^\mu dx^\nu - dz^2
\right\}, \qquad z \in \left[R, R'\right],
\end{equation}
where $k$ is the curvature scale, $R = \frac{1}{k}$, $R' =
\frac{e^{kL}}{k}$, and $L$ is the (proper) length of the fifth dimension
which we choose to be $\sim \frac{35}{k}$ to explain the Planck-weak
hierarchy.
%
%
The Standard Model (SM) gauge group $SU(3)_C \times
SU(2)_L\times U(1)_Y$ is a subgroup of the bulk
gauge symmetry. To be specific, we take
the bulk gauge symmetry to be
$
%
%
SU(3)_C\times SO(5)\times U(1)_X$ in the following analysis (the
group algebra of $SO(5)$ can be found in Appendix \ref{SO5}). We
will drop the color group $SU(3)_C$ in the following analysis since
it does not affect our result. The gauge boson action is given by
\begin{eqnarray}\label{gaugekin}
S_g  = \int d^5 x \sqrt{-G}\left[
-\frac{1}{2g_5^2}\text{Tr}(F^{(A)MN} F^{(A)}_{MN}) -
\frac{1}{4g_X^2}F^{(X)MN}F^{(X)}_{MN} \right],
\end{eqnarray}
with
\begin{equation}
A_M = \sum_{a=1}^3 A_M^{a_L} T^a_L + \sum_{a=1}^3 A_M^{a_R} T^a_R +
\sum_{ \hat{a} =1}^3 A_M^{\hat{a}} T^{\hat{a}} + A_M^{\hat{4}} T^{\hat{4}},
\end{equation}
where $T^{a}_{L,R}$ are the generators of $SU(2)_L\times SU(2)_R
\cong SO(4) \subset SO(5)$, and $T^{\hat{a},\hat{4}}$ are the
generators of the coset $SO(5)/SO(4)$. $X_M$ is the gauge boson of
$U(1)_{X}$. The boundary conditions are chosen such that only the
subgroup $SU(2)_L \times U(1)_Y$ is unbroken at UV brane ($z = R$)
and $SU(2)_L \times SU(2)_R\times U(1)_X \cong SO(4)\times U(1)_X$
is unbroken at IR brane ($z = R'$), where the hypercharge Y is
defined as $\frac{Y}{2} = T^{3}_R + Q_X$. Specifically, we choose
the $A_\mu \,(\mu = 0,1,2,3)$ components of $SU(2)_L \times U(1)_Y$
and $SO(4)\times U(1)_Y$ to have Neumann boundary condition
(``$+$'') on the UV brane and IR brane respectively, and all the
other $A_\mu \,(\mu = 0,1,2,3)$ have Dirichlet boundary condition
(``$-$'') on both branes. To reproduce hypercharge in Standard
Model, we do the following rotation of fields \cite{Medina:2007hz}
\begin{eqnarray}
\left(\begin{array}{c}A'^{3_R}_M \\ B^Y_M \end{array}\right) =
\left(
\begin{array}{cc} c_\phi & -s_\phi\\ s_\phi &
c_\phi\end{array}\right)\left(\begin{array}{c} A^{3_R}_M
\\ X_M\end{array}\right),\\
c_\phi = \frac{g_5}{\sqrt{g_5^2 + g_X^2}}, \qquad s_\phi =
\frac{g_X}{\sqrt{g_5^2 + g_X^2}},
\end{eqnarray}
where we need $s_\phi^2 \approx \tan^2\theta_W \approx 0.30$ to get
the correct Weinberg angle. Based on this definition, we set
$B^Y_\mu$ to have ``$+$'' boundary condition on both branes, and
$A'^{3_R}$ to have ``$-$'' boundary condition on UV brane and
``$+$''boundary condition on IR brane. With this set of assignment
of boundary conditions, we can reproduce SM gauge group at low
energy, while at the same time preserve $SU(2)_L\times SU(2)_R$
custodial symmetry \cite{Agashe:2003zs}.\\
An important observation here is that for gauge fields $A_M$, its
$A_\mu$ and $A_z$ components should have opposite boundary
conditions on the branes. This means that $A_z^{\hat{a}, \hat{4}}$
have ``$+$'' boundary conditions on both branes, thus there are zero
modes associated with them. We identify these zero modes of
$A_z^{\hat{a}, \hat{4}}$ as the Higgs fields $H^{a,4}$. They
transform as a doublet under $SU(2)_L$, thus have the same gauge
quantum numbers of SM Higgs. Due to 5D gauge invariance, these Higgs
fields are massless at tree level, and their potential is generated
by the breaking of $SO(5)$ on UV and IR branes. Therefore, the Higgs
potential will be generated through loop effects. Since from 5D
point of view, this is a non-local effect, the generated Higgs
potential will be finite. We will discuss the mechanism of radiative
generation of Higgs potential later in this section.

\subsection{Fermions}

The fermions also propagate in the bulk, with the following action
\begin{eqnarray}
S_f = \int d^5 x \sqrt{-G} \sum_i \bar{\Psi_i}(i \Gamma^M D_M - c_i
k) \Psi_i,
\end{eqnarray}
where $D_M = \partial_M - i A_M - i X_M$ and $c_i$ are the bulk
masses of the 5D fermions in units of $k$, which control the localization
of fermion zero modes. To be specific, the zero modes for left-handed (right-handed)
fermions are localized near UV brane if $c > 1/2 ~(c<-1/2)$, and they are
localized near IR brane if $c< 1/2 ~(c>-1/2)$. For future use, we define
\begin{eqnarray}\label{Eq. fc}
f(c) = \sqrt{\frac{1/2-c}{1-e^{-(1-2c)kL}}}~,
\end{eqnarray}
which is the size of zero mode fermion wavefunction at IR brane in
units of $\sqrt{2k}$. There are various scenarios to embed SM
fermions into representations of
$SO(5)$\cite{Medina:2007hz,Agashe:2004rs,Csaki:2008zd}. For the
following discussion, we just consider the third generation
fermions, since the first two generation fermions are not important
for EWSB and collider phenomenology. For concreteness, we follow
\cite{Medina:2007hz} and choose the fermion representation to be
$5\oplus 5\oplus 10$ for one generation. The generators of $SO(5)$
for $\mathbf{5}$ representation can be found in Appendix \ref{SO5}.
The fermions in $\mathbf{5}$ of $SO(5)$ have the following charge
assignment under $SU(2)_L \times SU(2)_R$:
\begin{equation}
\mathbf{5} = \frac{1}{\sqrt{2}}\left(\begin{array}{c} i q_{++} +
iq_{--}
\\ q_{--} -  q_{++}\\ iq_{-+} - iq_{+-}\\ q_{-+}+ q_{+-} \\ \sqrt{2}q^c
\end{array}\right),
\end{equation}
where $\pm$ means $\pm 1/2$ under $SU(2)_L$ and $SU(2)_R$
respectively, and $q^c$ means singlet. A more convenient basis is
\begin{equation}\label{Eq. 5rep}
\xi_5 =\left(\begin{array}{c} q_{++}
\\ q_{-+}\\ q_{+-}\\ q_{--} \\ q^c \end{array}\right) \equiv \left(\begin{array}{c}
\chi
\\ \tilde{t}\\t \\ b \\ \hat{t} \end{array}\right),
\end{equation}
where $\chi, t, b$ denote fermions with charge $+5/3, +2/3, -1/3$
respectively. The transformation between the two basis is
\begin{equation}\label{Eq:A}
\xi_5 = A\times \mathbf{5} \quad \text{with}\quad A =
\frac{1}{\sqrt{2}}\left(\begin{array}{ccccc} -i & -1 & 0 & 0 & 0\\ 0
& 0 & -i & 1 & 0\\ 0 & 0 & i & 1 & 0 \\ -i & 1 & 0 & 0 & 0 \\ 0 & 0
& 0 & 0 & \sqrt{2}
\end{array}\right) .
\end{equation}
The fermions are embedded in $\mathbf{10}$ of $SO(5)$ as follows
\begin{eqnarray}
\mathbf{10} = \left(\begin{array}{cccccccccc} \chi & \tilde{t} & t &
b & \Xi' & T' & B' & \Xi & T & B
\end{array} \right)^T ,
\end{eqnarray}
where $
 \left( \begin{array}{cc} \chi & t\\
\tilde{t} & b
\end{array}\right)
$ form an $SU(2)_L\times SU(2)_R$ bidoublet, $(\Xi, T, B)$ form
$SU(2)_R$ triplet, and $(\Xi', T', B')$ form $SU(2)_L$ triplet. We
can also write down the $\mathbf{10}$ representation in the form of
$5\times 5$ matrix
\begin{eqnarray}\label{Eq 5by5}
\xi_{10} = \frac{1}{2}\left( \begin{array}{ccccc}0 & T' + T & i
\frac{B'-\Xi'}{\sqrt{2}} + i \frac{B-\Xi}{\sqrt{2}} & \frac{B'+
\Xi'}{\sqrt{2}} - \frac{B + \Xi}{\sqrt{2}} & b + \chi\\ - T'- T & 0
& \frac{B' + \Xi'}{\sqrt{2}}+ \frac{B + \Xi}{\sqrt{2}} & - i\frac{B'
- \Xi'}{\sqrt{2}}+i \frac{B-\Xi}{\sqrt{2}} & i(b-\chi)\\-i
\frac{B'-\Xi'}{\sqrt{2}} - i \frac{B-\Xi}{\sqrt{2}} & -\frac{B' +
\Xi'}{\sqrt{2}}- \frac{B + \Xi}{\sqrt{2}} & 0 & T'- T & t +
\tilde{t}\\-\frac{B'+ \Xi'}{\sqrt{2}} + \frac{B + \Xi}{\sqrt{2}}& i
\frac{B'-\Xi'}{\sqrt{2}}-i\frac{B-\Xi}{\sqrt{2}} & -T' + T & 0 &
-i(t-\tilde{t})\\ - b-\chi & -i(b-\chi) & -t - \tilde{t}&
i(t-\tilde{t}) & 0
\end{array}\right) .
\end{eqnarray}
The fermion content and the boundary condition assignment can be
summarized as follows
\begin{eqnarray}
\Psi_{1L} =\left(\begin{array}{c} \chi_{1L} (-,+)
\\ \tilde{t}_{1L} (-,+)\\t_{1L} (+,+) \\ b_{1L} (+,+) \\ \hat{t}_{1L} (-,+) \end{array}\right)
,\qquad \Psi_{2R} =\left(\begin{array}{c} \chi_{2R} (-,+)
\\ \tilde{t}_{2R} (-,+)\\t_{2R} (-,+) \\ b_{2R} (-,+) \\ \hat{t}_{2R} (+,+)
\end{array}\right), \quad
\Psi_{3R} = \left(\begin{array}{c} \chi_{3R} (-,+)\\ \tilde{t}_{3R} (-,+)\\ t_{3R} (-,+)\\ b_{3R} (-,+)\\ \Xi'_{3R} (-,+) \\
T'_{3R} (-,+) \\ B'_{3R}(-,+) \\ \Xi_{3R}(-,+) \\ T_{3R}(-,+)\\
B_{3R} (+,+)
\end{array} \right),
\end{eqnarray}
while the opposite chirality fields have the opposite boundary
conditions. From this set of boundary conditions, we can see that
there are fermion zero modes for one $SU(2)_L$ doublet and two
$SU(2)_L$ singlets, which reproduce the SM fermion gauge
representations at low energy. To get SM fermion masses, we need the
following boundary mass terms
\begin{eqnarray}\label{fermion boundary terms}
{S}_{b} = \int d^5 x \sqrt{-G}\,\, 2 (k z)\delta(z- R')\left[M_{B_1}
\bar{\hat{t}}_{1L} \hat{t}_{2R} + M_{B_2} (\bar{\chi}_{1L},
\bar{\tilde{t}}_{1L}, \bar{t}_{1L}, \bar{b}_{1L})\left(\begin{array}{c} \chi_{3R}\\
\tilde{t}_{3R}\\ t_{3R}\\ b_{3R}\end{array} \right) +
\text{h.c.}\right] .
\end{eqnarray}
We have to choose the parameters $c_1, c_2, c_3, M_{B_1}, M_{B_2}$
to reproduce the top and bottom masses.

\subsection{Higgs Potential and KK Decomposition}

We have identified Higgs fields as the 5th components of the gauge
fields of coset $SO(5)/SO(4)$. Here, we briefly review the KK
decomposition of bulk fields with a background Higgs fields and how
the potential of Higgs is radiatively generated . For more details,
see
\cite{Medina:2007hz,Falkowski:2006vi}.\\

We denote $A_\mu^a$ as the gauge bosons of $SU(2)_L\times SU(2)_R$
and $A_\mu^{\hat{a}}$ ($\hat{a} = 1...4$)
as the gauge bosons of $SO(5)/SO(4)$. The zero
mode of $A_z^{\hat{a}}$ gives the Higgs. We can do the following KK
decomposition
\begin{eqnarray}
A_\mu^a(x,z) &=& \sum_n f^a_n(z,v) A_{\mu,n}(x),\\
\nonumber A_5^a(x,z) &=& \sum_n \frac{\partial_z f^a_n(z,v)}{m_n(v)}
h_{n}(x),\\ \nonumber A_\mu^{\hat{a}}(x,z) &=& \sum_n
f_n^{\hat{a}}(z,v) A_{\mu,n}(x),\\ \nonumber A_5^{\hat{a}}(x,z) &=&
{C_h} h^{\hat{a}}(x){kz} + \sum_n \frac{\partial_z
f^{\hat{a}}_n(z,v)}{m_n(v)} h_n(x).
\end{eqnarray}
We need $C_h = \sqrt{\frac{2k}{ (e^{2kL} -1)}} g_5$ to make the
Higgs field canonically normalized. Note that all the wavefunctions
depend on the vev of Higgs ($\langle h^{\hat{4}}\rangle = v$). The
boundary conditions for these wavefunctions are complicated.
However, the wavefunctions with non-vanishing Higgs vev are related
to the wavefunctions with vanishing Higgs vev by a gauge
transformation \cite{Falkowski:2006vi}
\begin{equation}\label{Eq trans}
f^\alpha(z,v) T^\alpha = \Omega^{-1}(z,v) f^\alpha(z,0)
T^\alpha\Omega(z,v),
\end{equation}
with
\begin{equation}\label{Eq. gauge transformation}
\Omega(z,v) = e^{-i C_h v T^{\hat{4}} \int_R^z dz' kz'} =
exp\left[-i C_h v T^{\hat{4}}  k(z^2-R^2)/2\right] \equiv exp\left[
-i \frac{v(z^2-R^2)}{f_h(R'^2-R^2)}T^{\hat{4}}\right ]
\end{equation}
where we defined the ``Higgs decay constant'' $f_h \equiv
\frac{\sqrt{2k}}{g_5\sqrt{e^{2kL}-1}}$. Therefore, to simplify the
task, we can just calculate the wavefunctions with vanishing Higgs
vev, and do a transformation $\Omega(z,v)$ to find the wavefunctions
with non-vanishing Higgs vev. We then apply boundary conditions for
the wavefunctions $f^{a,\hat{a}}(z,v)$ on the IR brane to get the
mass spectrum of gauge KK modes. The details of the calculation are
shown in Appendix \ref{KK decomp}. In the end, we get two spectral
functions $\rho_{W,Z}(m, v)$ for $W,Z$ bosons (Eqs.~(\ref{W spec})
and (\ref{Z spec})), whose roots give us
the mass spectra $m_{W,Z}^n$ for $W,Z$ bosons. \\
Similarly, the wavefunctions for fermions with non-vanishing Higgs
vev $F^\Psi_{1,2,3}(z, v)$ are also related to the wavefunctions for
fermions with vanishing Higgs vev $F^\Psi_{1,2,3}(z,0)$ by the gauge
transformation $\Omega(z,v)$:
\begin{eqnarray}\label{fermion gauge transformation}
F^\Psi_{1,2}(z,v) &=& A\Omega(z,v)^{-1}A^{-1} F^{\Psi}_{1,2}(z,0)\\
\nonumber F^{\Psi}_3(z,v) &=& \Omega(z,v)^{-1}
F^{\Psi}_3(z,0)\Omega(z,v)
\end{eqnarray}
where we have written $F_{1,2}^{\Psi}$ in the basis specified in Eq.
(\ref{Eq. 5rep}) and $F_3^{\Psi}$ in the form of $5\times 5$ matrix
(see Eq. (\ref{Eq 5by5})), and matrix $A$ is defined in Eq. (\ref{Eq:A}). Similarly to the gauge boson case, we can
get spectral functions for top and bottom quarks $\rho_{t,b}(m,v)$
(Eqs.~(\ref{bottom spec}) and (\ref{top spec})), whose roots give us
the mass spectra $m_{t,b}^n$ for $t,b$ fermions. We can calculate
the Coleman-Weinberg potential for Higgs once we know all the
spectral functions \cite{Falkowski:2006vi}
\begin{equation}\label{Higgs potential}
V_{CW}^W(v) = \frac{1}{(4\pi)^2} \int_0^\infty dp\,\, p^3
\left\{6\ln[{\rho}_W(ip,v)] + 3\ln[{\rho}_Z(ip,v)]
-12\ln[\rho_t(ip,v)] -12 \ln[\rho_b(ip,v)] \right\}.
\end{equation}
This integral can be done numerically. We can minimize this
potential to find the Higgs vev $v$. Then we can find the mass
spectra of the model through the spectral functions
$\rho_{W,Z,t,b}(m,v)$.

\subsection{Couplings of Coset Gauge Bosons}\label{couplings}

\subsubsection{Estimates and General Patterns}
\label{estimate}

The exact couplings for coset gauge bosons involve overlap integrals
of wavefunctions, which has to be done numerically and hence are not
very illuminating. We defer showing the formulae for exact couplings
to section \ref{exact} and a discussion of the numerical analysis to
section \ref{numerical}. In order to gain some insights into the
structure of coset gauge boson couplings, we concentrate here on
estimating the sizes of the couplings between both charged ($W_C$)
and neutral ($Z_C$) coset gauge bosons and fermions based on 5D
profiles, and we will show that the results here match the ones
coming from two-site description shown earlier in section
\ref{sec:two-site}.\footnote{The coupling between coset gauge
bosons and two SM gauge bosons are not studied here since they
vanish at leading order in Higgs vev due to quantum number (as
argued in Section \ref{sec:two-site}) and thus are not relevant for
collider study.} In the following analysis, we focus on the
parametric dependence of these couplings on $\theta_H \equiv
\frac{h}{\sqrt{2}f_h}$ and wavefunctions of fermion zero modes,
both of which give rise to more than an order-of-magnitude effect on the couplings.
There are also effects coming from fermion boundary mixing terms
(Eq.~(\ref{fermion boundary terms})), which will introduce only
order one uncertainty in our estimates.
%
%
However, the dependence of the couplings
on the parameters $\theta_H$ and
wavefunctions of fermion zero modes should be robust against
the effects from these mixing terms.

A comment is in order here about the region of parameter space we
are considering. As pointed out in \cite{Contino:2006qr, Medina:2007hz}, a light $t^{(1)}$ (first KK mode of top quark) is a
promising signature for GHU.  We will see later that a light
$t^{(1)}$ is also desirable for the collider study of the coset
gauge bosons. We generically get  {\em  two} light $t^{(1)}$ states in the regions of
parameter space when $c_1 < 0$.
In this case, the SM $(t,b)_L$ profile is highly peaked near
the TeV brane and thus
the SM $t_R$ is less so (in order to obtain the correct top quark mass).
We find that one of the light $t^{(1)}$ states is mostly $SU(2)_L$ {\em singlet} in this case.
Thus, the coupling of SM bottom (doublet) and this light $t^{ (1)}$ to the coset gauge boson
(doublet) is large since it is allowed by the quantum numbers (i.e., no need for EWSB)
and is not suppressed by profiles either. This coupling can then give a significant contribution to the production of the coset gauge boson.
Therefore, we focus on
this region of parameter space. We will often denote this singlet light $t^{ (1)}$
as ``{\em the} light $t^{ (1)}$'' in what follows. To simplify notation, we will also use $t, b$
to denote SM top and bottom fermions when there is no confusion.

References \cite{Carena:2006bn} showed that the one loop contributions of such light $t^{(1)}$ states
to the $T$ parameter and to the shift in $Z b \bar{b}$ coupling can be consistent with the
data.
Another potential constraint comes from the shift in the $Wtb$ coupling.
We have numerically studied the shift in the $tbW$ coupling induced
by mixing of zero and KK modes of both $W$ and top (including
the effect of the light $t^{(1)}$ state).
We find that
this shift is smaller than
$\sim 10 \%$, as required by the recent measurements at Tevatron \cite{Group:2009qk}.
Alternatively, the SM $t_R$ can be highly peaked near the TeV brane [and the SM $(t,b)_L$ less so], which results in the light $t^{ (1) }$ being a doublet \cite{Contino:2006qr}
and a large coset-$t_R$-$t^{ (1) }$
coupling. However, the top quark content of the proton is negligible
(cf. bottom quark content which is larger) so that this coupling
will not be that useful for production of coset gauge bosons.

\begin{itemize}

\item

\underline{Charged Coset Gauge Bosons ($W_C$)}

\item $g_{W_C t b}$: coupling between coset $W_C$, SM top
and SM bottom. We first discuss the coupling for left-handed
fermions. Once the Higgs boson gets a vev, there will be mixing between $W_L$
and $W_C$ and between $t_{1L}$ and $\hat{t}_{1L}$. From another
point of view, this mixing comes from the gauge transformation
(Eqs.~(\ref{Eq trans}) and (\ref{fermion gauge transformation}))
that link the wavefunctions with vanishing Higgs vev and
wavefunctions with non-vanishing Higgs vev. For example, from
Eq.~(\ref{Eq gauge functions with nonvanishing vev}) we can see the
wavefunction of $W_L$ with non-vanishing Higgs vev contains some
part of $W_C$ wavefunction with vanishing Higgs vev, and the amount
is $\frac{\sin\theta_H}{\sqrt{2}}$. Therefore the dominant
contribution to the coupling comes from the following overlap
integral of wavefunctions
\begin{equation}
g_{W_C t_L b_L} \approx -\frac{g_5}{\sqrt{2}}\int
\frac{dz}{z^4}\left[ f_{W_L} F_{{t}_1} F_{b_1} + f_{W_C}
F_{{\hat{t}}_1} F_{b_1} \right]
\end{equation}
The first term comes from the mixing between $W_L$ and $W_C$, and
the second term comes from the mixing between $t_{1L}$ and
$\hat{t}_{1L}$. Here we have assumed that the zero mode $t_L, b_L$ lives
mainly in the first fermion multiplet $\Psi_1$. This happens when
$c_1 < 0$.  To estimate this coupling, we need to know the
wavefunctions of $t_L$, $b_L$ and $W_C$. The wavefunctions of $W_C$
are peaked near the IR brane (i.e., their size at the IR brane is
$\sim O(1)$
in units of $\sqrt{k}$) since they are KK modes, and {\em each} of the
wavefunctions of $t_L$ and $b_L$ at the IR brane is $f(c_1)\approx
\sqrt{\frac{1}{2}-c_1}$ (in units of $\sqrt{k}$). Finally,
the overlap integral will be dominated by
a region of size $\sim \frac{1}{k}$ near the IR brane. This gives us
an estimate
\begin{equation}\label{Wctbest}
|g_{W_C t_Lb_L}| \sim \frac{g_5 \sqrt{k}}{\sqrt{2}}
\left(\frac{1}{2}-c_1\right) \frac{\sin\theta_H}{\sqrt{2}},
\end{equation}
where $\frac{\sin\theta_H}{\sqrt{2}}$ comes from mixing induced by
Higgs vev. From Eq.~(\ref{Wctbest}) we can see that it is possible
to get order one coupling between $W_C$ and SM top and bottom
quarks.\footnote{Clearly, the coupling
analogous to Eq.~(\ref{Wctbest}) is
negligible for light left/right-handed
SM fermions
which have $c > (<) 1/2 (-1/2)$ and hence $f(c) \ll 1$.
In particular, the coset gauge boson wavefunctions vanish near the
Planck brane so that the wavefunction overlap comes only from near the
TeV brane, unlike for KK $W/Z$ where the
flavor-universal part of the coupling to two SM fermions
comes from overlap near the Planck brane.}
The right-handed coupling $g_{W_C t_R b_R}$ should be much smaller
than the left-handed coupling $g_{W_C t_Lb_L}$ since the
wavefunction of $b_R$ is much smaller than that of $b_L$ near the IR
brane. Therefore, it is irrelevant for collider study.

\item $g_{W_C t^{(1)}b}$: coupling between coset $W_C$, first top KK mode, SM
bottom quark. We first study the left-handed coupling.
Note that the
$t^{(1)}$ is mostly $SU(2)_L$ singlet and its wavefunction is also peaked near the IR brane (i.e.,
its size at
the IR brane is $\sim O(1)$ in units of $\sqrt{k}$), just like
$W_C$.
Thus the size of this coupling should be controlled simply by the single
$b$ wavefunction near the IR brane (i.e., no factor of EWSB). Therefore the coupling should be
of order $\frac{g_5\sqrt{k}}{\sqrt{2}} f(c_1) \approx
\frac{g_5\sqrt{k}}{\sqrt{2}}\sqrt{\frac{1}{2}-c_1}$ in the $\theta_H
\to 0$ limit. Including the effect of nonzero Higgs vev will only
give a small correction to this coupling, thus the estimate remains
the same. The coupling for right-handed fermions should be much
smaller due to the same reason that the wavefunction of $b_R$ near
the IR brane is small.

\item $g_{W_C t^{(1)} b^{(1)}}$: coupling between coset $W_C$, first top KK mode
and first bottom KK mode. Since the KK modes of fermions are
localized near the IR brane, the coupling for both left-handed and
right-handed fermions should be of order
$\frac{g_5\sqrt{k}}{\sqrt{2}}$ (i.e., no suppression due to profiles or EWSB), up to order one coefficients coming
from boundary mixing terms.

\item

\underline{Neutral Coset Gauge Boson ($Z_C$)}

\item $g_{Z_C t t}$: coupling between neutral coset gauge boson
 $Z_C$ and SM top
quark. For left-handed coupling, the estimate is similar to that of
$g_{W_C t_L b_L}$:
\begin{equation}
g_{Z_C  t_L t_L} \sim
{g_5\sqrt{k}}\left(\frac{1}{2}-c_1\right)\frac{\sin(\theta_H)}{\sqrt{2}}
\end{equation}
For right-handed coupling, the estimate is also similar
\begin{equation}
g_{Z_C  t_R t_R} \sim
{g_5\sqrt{k}}\left(\frac{1}{2}+c_2\right)\frac{\sin(\theta_H)}{\sqrt{2}}
\end{equation}

\item $g_{Z_C t^{(1)} t}$: coupling between coset $Z_C$, KK top and
SM top quark. For left-handed coupling the estimate is
\begin{equation}
 g_{Z_C t^{(1)}_L t_L} \sim {g_5\sqrt{k}}\sqrt{\frac{1}{2}-c_1}
\end{equation}
and for right-handed coupling
\begin{equation}
 g_{Z_C t^{(1)}_R t_R} \sim {g_5\sqrt{k}}\sqrt{\frac{1}{2}+c_2}
\end{equation}

\item $g_{Z_C t^{(1)} t^{(1)}}$: coupling between coset $Z_C$ and KK top quark.
Since  $Z_C$ always couples to two fermions transforming in different
representation of $SU(2)_L$, this coupling will not be generated in
the $\theta_H \to 0$ limit. Therefore, a rough estimate of this
coupling is
\begin{equation}
g_{Z_C t^{(1)} t^{(1)}} \sim g_5\sqrt{k}\frac{\sin
\theta_H}{\sqrt{2}}
\end{equation}
This estimate holds for both left-handed and right-handed couplings
since the wavefunctions for $t^{(1)}_L$ and $t^{(1)}_R$ are both IR
localized.

\item  $g_{Z_C b b}$: coupling between neutral coset gauge boson
$Z_C$ and SM bottom quark. For the left-handed coupling, naive
estimate will give us
\begin{equation}
g_{Z_C  b_L b_L}^{\text{naive}} \sim
{g_5\sqrt{k}}\left(\frac{1}{2}-c_1\right)\frac{\sin\theta_H}{\sqrt{2}}
\end{equation}
However, this coupling is very small (i.e.,
not relevant for collider signals) due to custodial
symmetry. To be specific, the two
contributions to this coupling coming from $Z_C$ mixing with
$W_{L,R}^3$ cancel each other. Similarly,
the contributions to this coupling coming from $(2,2)$
fermion mixing with $(1,3)$ and $(3,1)$ fermion cancel each other.
This cancellation is related to the build-in custodial symmetry of
the model that protects the $g_{Z b_L b_L}$ coupling (see Appendix
\ref{sec:zbb} for more detailed discussion). Note that this
cancelation does not happen for top quark since its $W_L^3$ and
$W_R^3$ charges are different. The right-handed coupling is small
due to the small $b_R$ wavefunction near IR brane.

\item  $g_{Z_C b b^{(1)}}$: coupling between coset $Z_C$,
SM bottom and KK bottom quark. The left-handed coupling is small due
to similar cancelation that suppress $g_{Z_C  b_L b_L}$ coupling.
The right-handed coupling is also small because of the small $b_R$
wavefunction near IR brane.

\end{itemize}

There is an additional coset gauge boson $A_\mu^{\hat{4}}$ (gauge
boson of the generator $T^{\hat{4}}$) which is the vector partner of
physical Higgs boson. We do not consider it here because its
coupling with two SM fermions vanishes.\footnote{The reason for this
is that Higgs vev does not induce an effect on $A_\mu^{\hat{4}}$
coupling since the gauge transformation (Eq.~(\ref{Eq. gauge
transformation})) commutes with $T^{\hat{4}}$.} Even though it has
nonzero coupling with $b b^{(1)}$ and $t t^{(1)}$, its production at
the LHC is still suppressed. The reason is that the $b^{(1)}$ is not
light (in the case of associated production with $b^{(1)}$ using the
coupling to $b b^{(1)}$) and the top quark content of the proton is
negligible, even though $t^{(1)}$ is light (in the case of
associated production with $t^{(1)}$ using the coupling to $t
t^{(1)}$).

We can compare the pattern of couplings estimated here with our
conclusion using the two-site approach. In fact, there is a
one-to-one correspondence/dictionary between two-site language and
warped extra dimension models (see \cite{Contino:2006nn,Agashe:2008uz}):
\begin{eqnarray}
\text{SM states} &\leftrightarrow& \text{zero modes}~, \\ \nonumber
\text{heavy states} &\leftrightarrow& \text{KK modes}~,\\ \nonumber
\sin\theta_{\psi_{L,R}} &\leftrightarrow& f(c_{L,R})~,\\ \nonumber
\sin\theta_G &\leftrightarrow& \frac{1}{\sqrt{kL}}~,\\ \nonumber g_*
&\leftrightarrow& g_5\sqrt{k}~.
\end{eqnarray}
Based on this identification, we can see that the estimates for
specific 5D
model agree with those obtained using two-site description, the latter
estimates
being applicable to the general warped/composite PGB Higgs
framework. Namely, the
coset gauge boson generally couple strongly with SM fermions (zero
modes) and heavy fermions (KK modes). We emphasize again that the
conclusions above for the 5D model are rough, but are quite general,
for example, they do not depend on whether the
bulk gauge symmetry breaking [$SO(5) \rightarrow SO(4)$] vev
is infinite
(as in GHU models) or finite\footnote{This insensitivity is related to
a similar one in the two-site description in section \ref{sec:two-site},
in the latter case
to
$f_{ \phi }$, the scalar vev breaking the global symmetry
[$SO(5) \rightarrow SO(4)$]
in the composite sector.}.
We will further validate these estimates by computing them
numerically for the specific 5D GHU model in section
\ref{numerical}.

\subsubsection{Exact Couplings}\label{exact}
The exact couplings of coset gauge bosons can be obtained by overlap
integrals of wavefunctions. We define gauge boson wavefunction
matrix
\begin{eqnarray}
G(z,v) \equiv f^{a_L}(z,v) T^{a_L} +f^{a_R}(z,v) T^{a_R} +
f^{{\hat{a}}}(z,v) T^{\hat{a}} + f^{{\hat{4}}}(z,v) T^{\hat{4}}~.
\end{eqnarray}
And we just use $F_{1,2,3}^\Psi(z,v)$ (see Eq.~(\ref{fermion gauge
transformation})) to denote the wavefunctions of the three fermionic
multiplets. Then we can get the coupling between fermions and coset
gauge bosons:
\begin{eqnarray}\label{coset coupling}
g_{GFF} =  \int  \frac{dz}{(k z)^4} \,\left\{g_5 \left[F_{1,2}^{\Psi
\dagger} A G A^\dagger F_{1,2}^{\Psi} + \text{Tr}\left(F_3^{\Psi
\dagger}[G, F_3^\Psi]\right)\right] + g_X \left[ f^X
\left(F_{1,2}^{\Psi\dagger} F_{1,2}^\Psi +
\text{Tr}\left[F_{3}^{\Psi\dagger} F_{3}^\Psi \right]\right)\right]
\right\} .
\end{eqnarray}
where matrix $A$ is defined in Eq. (\ref{Eq:A}). We use this formula to do numerical analysis in the next section.


\subsection{Numerical Results} \label{numerical}

The purpose of our numerical scan is to find some points in
the parameter space of the GHU model, in particular, specific
values of the coset gauge boson couplings, which can then
be used as benchmarks for
our collider study. Therefore, we pick a specific model of GHU
introduced in \cite{Medina:2007hz} to do our numerical scan.
However, we should {\em not} treat the results as being model dependent
for the following reason. As we
argued in previous sections, the masses of coset gauge bosons in
non-minimal models can be different from minimal GHU models.
However, the couplings of coset gauge bosons are not sensitive to
this non-minimal structure since their pattern is determined by the quantum
numbers and the fact that coset gauge boson wavefunctions are peaked
near the IR brane.

In our numerical scan, we fix $k = 10^{18}$ GeV and we scanned over
the input parameters $g_5\sqrt{k}, kL, c_{1,2,3}, M_{B1,B2}$. Even
though for minimal model we have $g_5\sqrt{k} \approx g\sqrt{kL}$,
this relationship between 5D coupling and 4D gauge coupling is
modified once we include brane kinetic terms for gauge fields.
Therefore, we choose to scan $g_5\sqrt{k}$ over the range
$[g\sqrt{kL}, 2g\sqrt{kL}]$. We calculate the Higgs potential (Eq.
(\ref{Higgs potential})) and minimize it to find the Higgs vev $v$.
We can then calculate the particle spectrum using the spectral
functions (Eqs.~(\ref{W spec}), (\ref{Z spec}), (\ref{bottom spec})
and (\ref{top spec})). We collect points with reasonable top and
$W/Z$ masses. Finally, the couplings of coset gauge bosons are
calculated using Eq.~(\ref{coset coupling}). The important couplings
are:
\begin{itemize}
\item[(i)] for charged coset gauge boson $W_C$
\begin{eqnarray}
{\cal L}_{W_C} &=& g_{W_C t_L b_L} \bar{t}_L \gamma_\mu b_L
W_C^{+\mu} + g_{W_C t_R b_R} \bar{t}_R \gamma_\mu b_R W_C^{+\mu}\\
\nonumber & +& g_{W_C t^{(1)}_L b_L} \bar{t}^{(1)}_L \gamma_\mu b_L
W_C^{+\mu}+ g_{W_C t^{(1)}_R b_R} \bar{t}_R^{(1)} \gamma_\mu b_R
W_C^{+\mu} + \text{h.c.} ,
\end{eqnarray}
\item[(ii)] for neutral coset gauge boson $Z_C$
\begin{eqnarray}
{\cal L}_{Z_C} &=& g_{Z_C t_L t_L}\bar{t}_L \gamma_\mu t_L Z_C^\mu +
g_{Z_C t_R t_R}\bar{t}_R \gamma_\mu t_R Z_C^\mu \\ \nonumber &+&
g_{Z_C t_L^{(1)} t_L}\bar{t}^{(1)}_L \gamma_\mu t_L Z_C^\mu + g_{Z_C
t_R^{(1)} t_R}\bar{t}^{(1)}_R \gamma_\mu t_R Z_C^\mu \\ \nonumber
&+& g_{Z_C b_L b_L}\bar{b}_L \gamma_\mu b_L Z_C^\mu + g_{Z_C b_R
b_R}\bar{b}_R \gamma_\mu b_R Z_C^\mu  \\ \nonumber &+& g_{Z_C
b^{(1)}_L b_L}\bar{b}^{(1)}_L \gamma_\mu b_L Z_C^\mu + g_{Z_C
b^{(1)}_R b_R}\bar{b}^{(1)}_R \gamma_\mu b_R Z_C^\mu + \text{h.c.} ,
\end{eqnarray}
\item[(iii)] for first KK top $t^{(1)}$
\begin{eqnarray}
{\cal L}_{t^{(1)}} &=& g_{Wt_L^{(1)} b_L}\bar{t}^{(1)}_L \gamma_\mu
b_L
W^{+\mu} +g_{Wt_R^{(1)} b_R}\bar{t}^{(1)}_R \gamma_\mu b_R W^{+\mu} \\
\nonumber &+& g_{Zt_L^{(1)} t_L}\bar{t}^{(1)}_L \gamma_\mu t_L Z^\mu
+ g_{Zt_R^{(1)} t_R}\bar{t}^{(1)}_R \gamma_\mu t_R Z^\mu \\
\nonumber &+& g_{Ht_L^{(1)} t_R}\bar{t}^{(1)}_L h t_R +
g_{Ht_R^{(1)} t_L}\bar{t}^{(1)}_R h t_L + \text{h.c.}  ,
\end{eqnarray}
\end{itemize}
where the subscripts $L$, $R$ imply the chirality of the
fermion. We present here a sample point with the couplings from the
scan in Table \ref{sampleinputpoints} and Table \ref{couplingtable}.
This will be served as benchmark point for collider study.

\begin{table}[hbt]
\begin{center}
\begin{tabular}{|c|c|c|c|c|c|c|c|}
\hline   $k e^{-kL}$ & $g_5 \sqrt{k}$ & $c_1$ &  $c_2$ & $c_3$ &  $M_{B1}$ & $M_{B2}$ & $\theta_H$ \\
\hline   $956\ \text{GeV}$ & 7.16 & -0.364 & -0.446 & -0.559 & 1.419&
-0.139 & 0.410 \\
\hline
\end{tabular}
\end{center}
\vspace{-.4cm} \caption[]{ Input parameters for sample points used
in our numerical calculation. We fix $k = 10^{18} ~\text{GeV}$. $c_{1,2,3}$ are the bulk mass parameters for the fermionic multiplets. $M_{B1}, M_{B2}$ are boundary mass parameters needed to get correct SM fermion masses (see Eq. \ref{fermion boundary terms}).
\label{sampleinputpoints}}
\end{table}
\begin{table}[hbt]
\begin{center}
\begin{tabular}{|c|c|c|c|c|}
\hline $g_{W_C t_L b_L}$ & $g_{W_C t_R b_R}$ & $g_{W_C t^{(1)}_L b_L}$ & $g_{W_C t^{(1)}_R b_R}$  \\
\hline 0.712 & 0.00169& -1.945 & 0.00207\\
\hline
\end{tabular}
\begin{tabular}{|c|c|c|c|c|c|c|c|c|}
 \hline $g_{Z_C t_L t_L}$ & $g_{Z_C t_R t_R}$ & $g_{Z_C
t^{(1)}_L t_L}$ & $g_{Z_C t^{(1)}_R t_R}$ & $g_{Z_C b_L b_L}$ &
$g_{Z_C b_R b_R}$ & $g_{Z_C b^{(1)}_L b_L}$ &
$g_{Z_C b^{(1)}_R b_R}$ \\
\hline  -0.930 & 0.119 & 1.242 & 0.177 & 0.0235 & -0.0219 & 0.0294 & 0.136 \\
\hline
\end{tabular}
\begin{tabular}{|c|c|c|c|c|c|}
\hline $g_{W t_L^{(1)} b_L}$ & $g_{W t_R^{(1)} b_R}$ & $g_{Z
t_L^{(1)} t_L}$ & $g_{Z
t_R^{(1)} t_R}$ & $g_{H t_L^{(1)} t_R}$ & $g_{H t_R^{(1)} t_L}$ \\
\hline  -0.170 & 0.000040 & 0.121 & -0.0888 & 0.654 & -1.06  \\
\hline
\end{tabular}
\end{center}
\vspace{-.4cm} \caption[]{ \label{couplingtable} Numerical values of the couplings for the  sample point choice.}
\end{table}


\section{LHC signals}
\label{signal}

As discussed in sections \ref{sec:two-site} and \ref{GHU}, the most
characteristic feature of coset gauge bosons is that they are vector
bosons possessing Higgs quantum number. This uniquely fixes the
pattern they are coupled to SM particles and other KK modes --- they
predominantly couple to one SM and one KK fermions as their
couplings to a pair of SM particles are only induced by EWSB and
are thus subdominant. This further determines how they are produced and decay
at the LHC.

As will be discussed in section \ref{production}, the production
rate of neutral coset KK modes at LHC is very low in general. We
will thus focus on the LHC signals of charged coset KK modes $W_C$.
Our study is based on a set of points in the parameter space that
give reasonable SM particle masses and generate EWSB radiatively as
discussed in section \ref{numerical}. We first discuss the decay
of charged coset gauge KK boson in \ref{decay} and its production at
the LHC in \ref{production} using this set of points. We then use
couplings corresponding to the representative benchmark point in
Table \ref{couplingtable} and discuss in detail the signal and
background at the LHC in \ref{subsec:signal}.

\subsection{Decay of $W_C$}
\label{decay}

\begin{figure}
\centerline{\epsfig{file=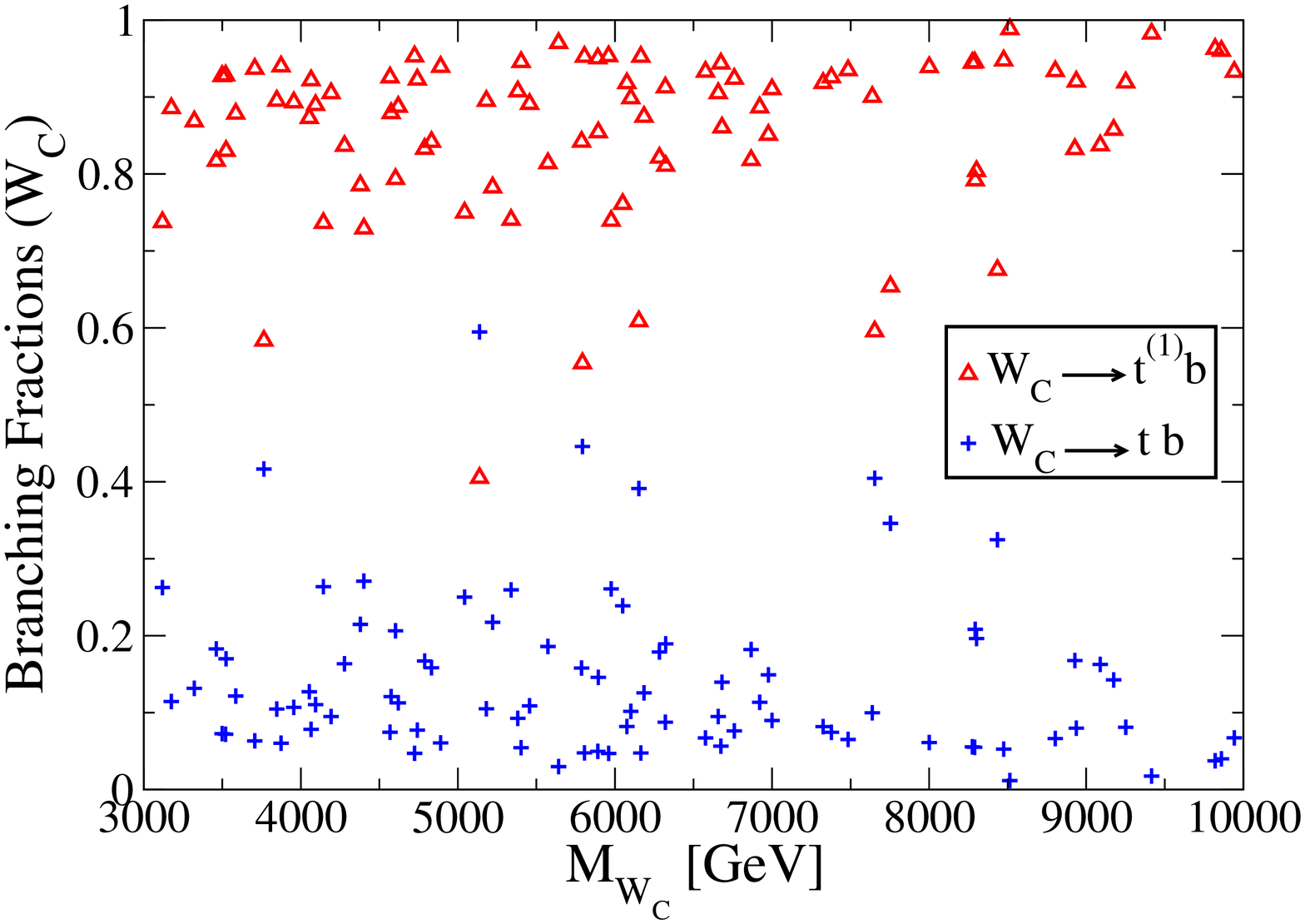,width=6.8cm,angle=-90}
\epsfig{file=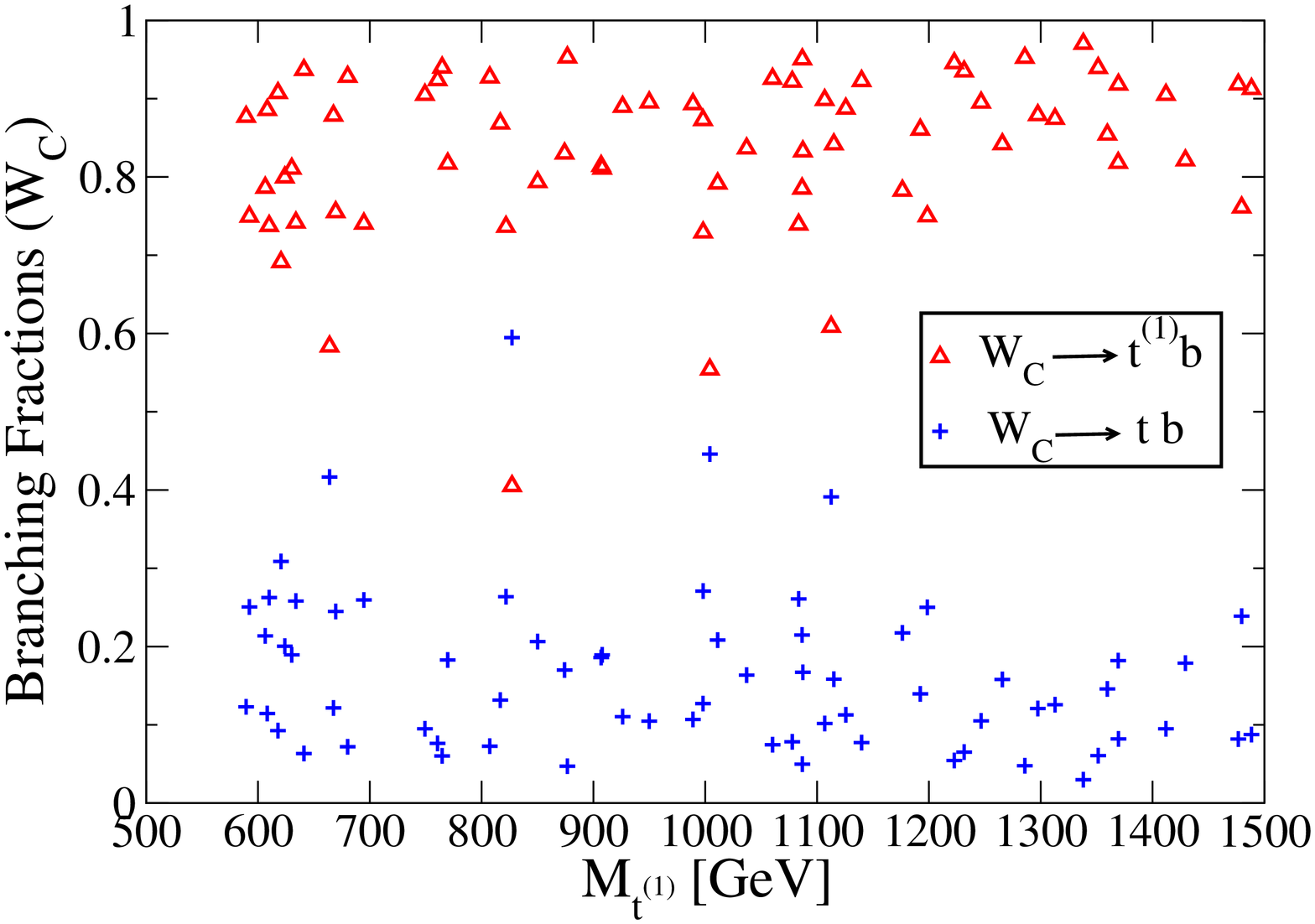,width=6.8cm,angle=-90} }
\caption{Scatter plots for the branching fractions Br$(W_C \rightarrow t^{(1)} b)$
(triangle symbol) and Br$(W_C \rightarrow t b)$ (cross symbol)
versus (a)  $M_{W_C}$ and (b) $M_{t^{(1)}}$, respectively.} \label{fig:wcdec}
\end{figure}

There are following decay channels of $W_C$\footnote{The other decay
channels involving light SM fermions are negligible.}
\begin{equation}
\label{eq:wcdec} W_C \rightarrow t b^{(1)},~ t^{(1)} b,~ t b.
\end{equation}
Compared to $W_C \rightarrow t^{(1)} b$, the branching fraction of $W_C
\rightarrow t b^{(1)}$ is substantially suppressed due to
kinematical reasons. It is one of the important properties of GHU
models that there exists a light KK mode of top quark $t^{(1)}$
\cite{Medina:2007hz}. The mass of $b$-quark KK mode $b^{(1)}$ is, on
the other hand, usually much heavier. Thus the decay $W_C
\rightarrow t b^{(1)}$ is in most cases highly suppressed, if not
forbidden. In the following, for simplicity and without losing the
general feature, we will assume $b^{(1)}$ is heavier than $W_C$,
forbidding this decay channel completely.

On the other hand, the branching fraction of $W_C \rightarrow t b$ is
much suppressed compared to $W_C \rightarrow t^{(1)} b$ due to
dynamical reasons. As discussed in sections \ref{sec:two-site} and
\ref{GHU}, the quantum number of $W_C$ forbids its coupling to SM
quarks like $\bar t b$
at leading order. This coupling is only induced
by Higgs vev after EWSB and is thus suppressed by $v/f_h$. This
determines the typical trend of branching fractions of $W_C \rightarrow
t^{(1)} ~ b$ and $W_C \rightarrow t ~ b$ decays, shown in Fig.
\ref{fig:wcdec}.

\begin{figure}
\centerline{\epsfig{file=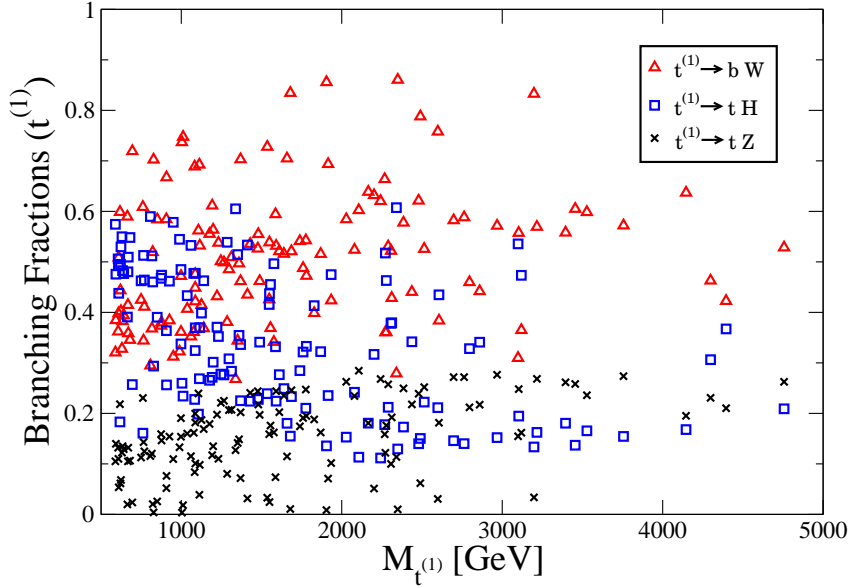,width=9.6cm,angle=-90}
}\caption{Scatter plots for the branching fractions Br$(t^{(1)} \rightarrow W b)$
(triangle symbol), Br$(t^{(1)} \rightarrow t h)$ (square symbol),
and Br$(t^{(1)} \rightarrow t Z)$ (cross symbol) versus $M_{t^{(1)}}$.} \label{fig:t1dec}
\end{figure}

Since $t^{(1)}$ appears in the most dominant decay channel of
$W_C$, we thus comment on $t^{(1)}$ decay next. There are three decay channels of $t^{(1)}$
\begin{equation}
\label{eq:t1decchns}
 t^{(1)} \rightarrow
bW,\ \ th, \ \ tZ.
\end{equation}
This has been studied in great detail in Ref.~\cite{Carena:2007tn},
where it has been pointed out that, for large $M_{t^{(1)}}$, the
branching fractions should follow the relation $2:1:1$, according to the
Goldstone Boson Equivalence Theorem, where as for small $M_{t^{(1)}}$,
this ratio does not hold. In Ref.~\cite{Carena:2007tn}, the
branching fractions are only shown for $M_{t^{(1)}}$ larger than 1 TeV.
Since our primary goal is to explore the reach of LHC on discovering
$W_C$, a relatively light $t^{(1)}$ would be more relevant. We thus
extend the $t^{(1)}$ decay branching fractions to a wider range of
500 GeV$-$5 TeV, as in Fig.~\ref{fig:t1dec}. The $2:1:1$
ratios hold for $M_{t^{(1)}} >$3 TeV. For a light $t^{(1)}$, $M_{t^{(1)}} < 1$ TeV,
the branching fractions of $t^{(1)}
\rightarrow bW$ and $t^{(1)} \rightarrow th$ are close and both
are significantly larger than that of $t^{(1)} \rightarrow tZ$. The
LHC search of $t^{(1)}$ has also been discussed in Ref.~\cite{Carena:2007tn}
and positive conclusions were reached. We therefore assume that $t^{(1)}$ has been
observed with its mass approximately known a priori to the searches for $W_C$.

\begin{figure}
\centerline{\epsfig{file=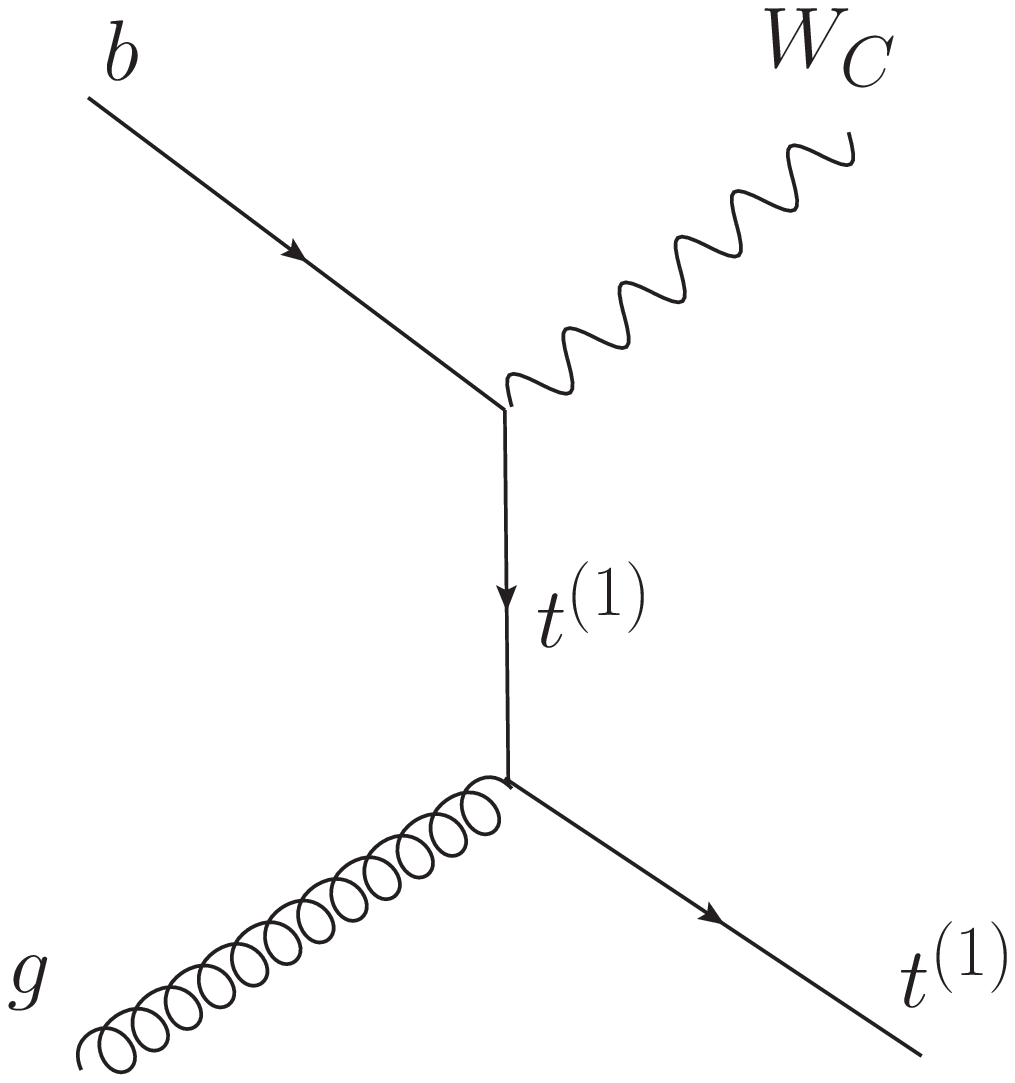,width=5cm,angle=0}
\epsfig{file=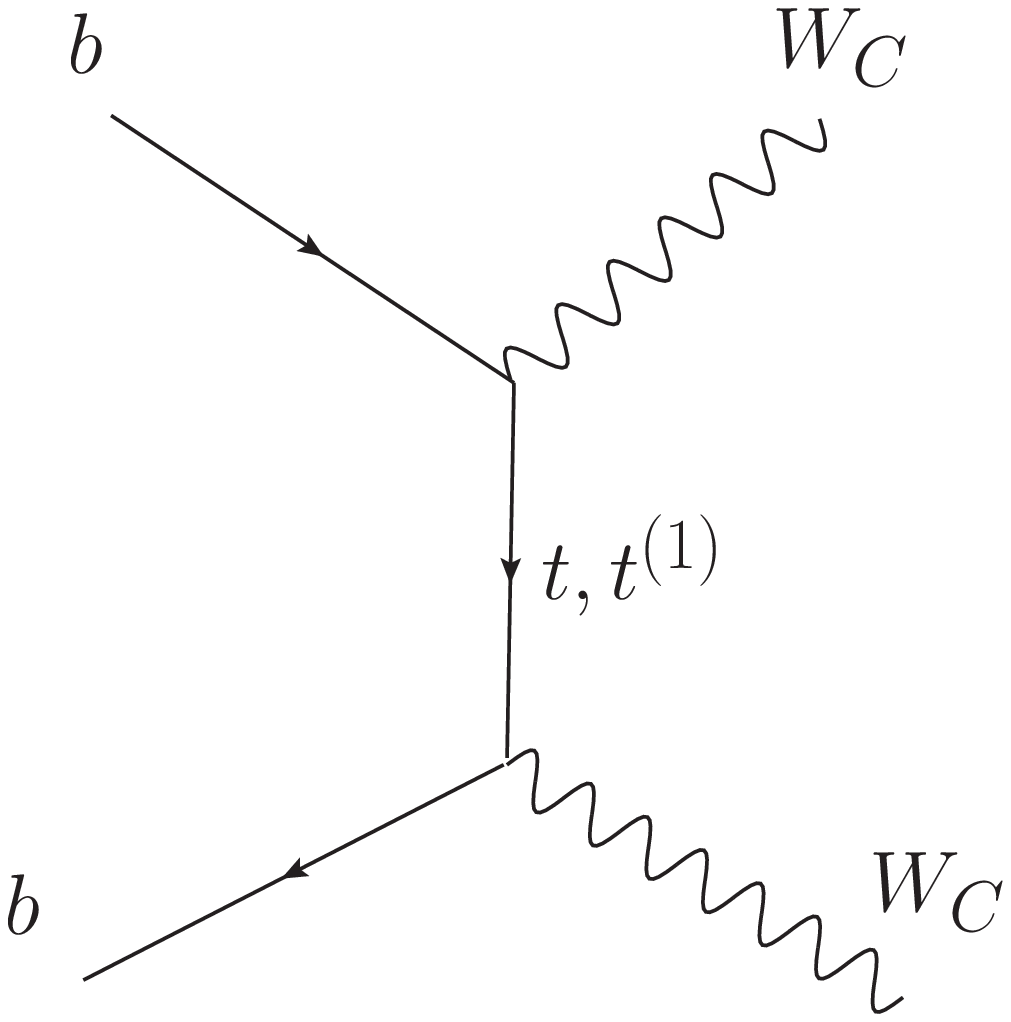,width=5cm,angle=0} \put(-225,0){(a)}
\put(-82,0){(b)} } \caption{Representative Feynman diagrams for (a) $W_C t^{(1)}$
associated production and (b) $W_C W_C$ pair production.}
\label{fig:Feynman1}
\end{figure}

\subsection{Coset KK modes production at the LHC}
\label{production}

\begin{figure}[tb]
\centerline{\epsfig{file=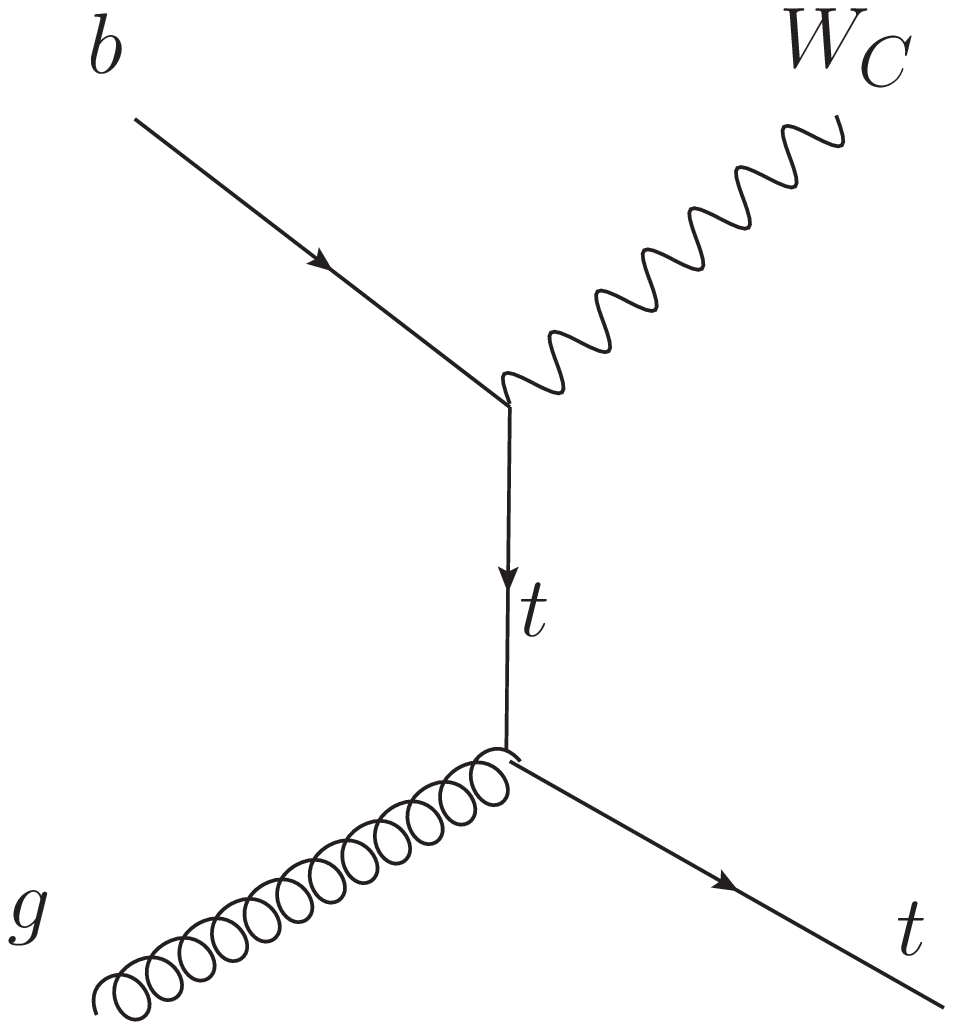,width=5cm,angle=0}
\epsfig{file=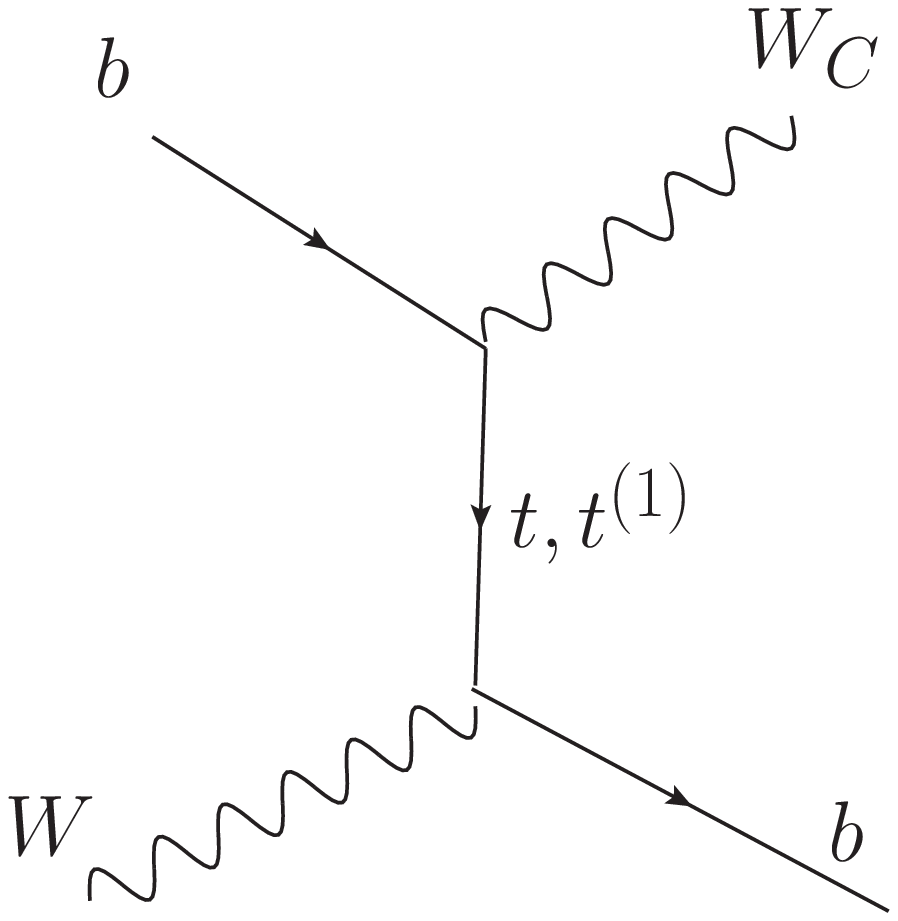,width=5cm,angle=0}
\epsfig{file=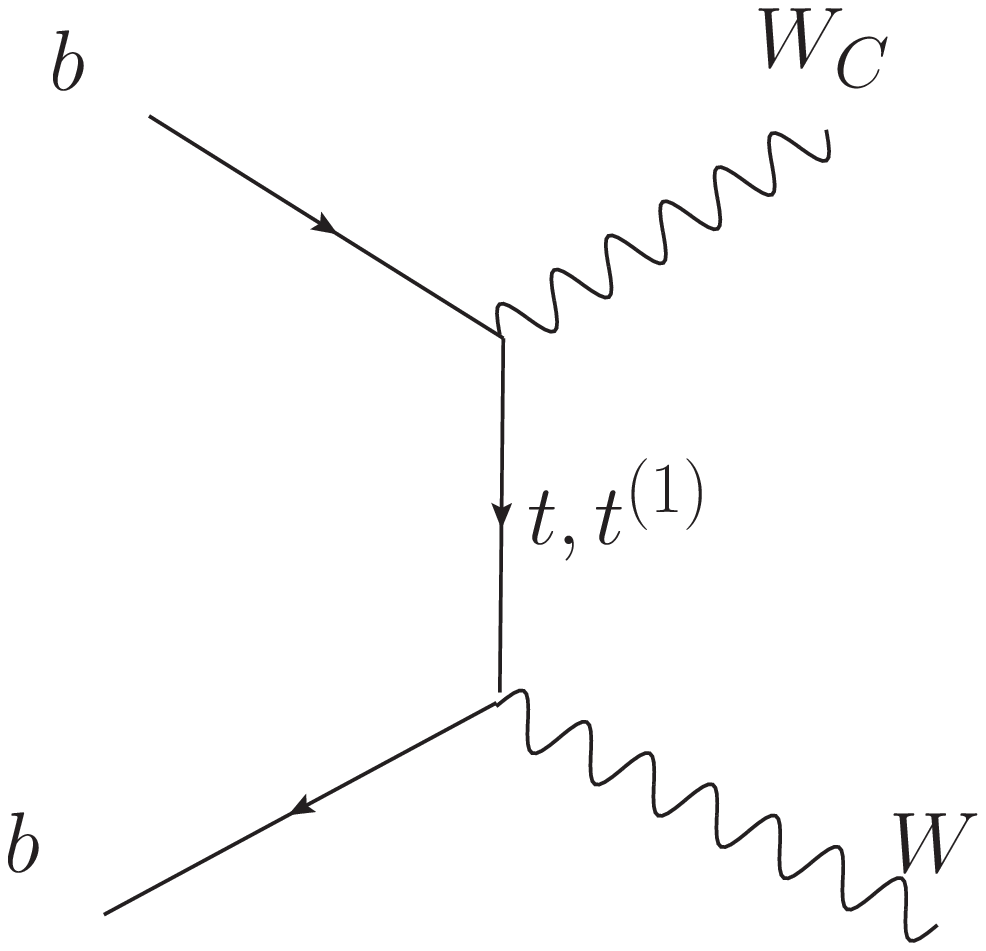,width=5cm,angle=0}\put(-370,0){(a)}
\put(-225,0){(b)}\put(-78,0){(c)}} \caption{Representative Feynman diagrams
for associated production (a) $W_C t$, (b) $W_C b$, and (c) $W_C W$, respectively.}
\label{fig:Feynman2}
\end{figure}

The coset gauge bosons, as all the KK modes in general, have
profiles with large overlap with the third generation SM fermions
($t,\ b$) and hence couple more strongly to them as compared to
the 1st/2nd generation SM quarks.
%
However,
%
%
the dominant production of KK $W/Z$ is still (typically)
via $u$ and $d$ quarks.
On the other hand, the coupling to light quarks is
smaller for the case of coset gauge bosons than the KK $W/Z$ (as
discussed in section \ref{sec:two-site}).
This feature
motivates consideration of coupling of coset gauge bosons to
bottom quarks
for their production at the LHC.
%
%
%
%
From the discussion in section
\ref{couplings}, and as shown explicitly in Table
\ref{couplingtable}, the neutral coset KK modes $Z_C$ couple
strongly to $t$($\bar{t}$), but rather weakly to $b$($\bar{b}$),
indicating that its production is highly suppressed at the LHC. We
will then focus on the production of charged ones $W^{\pm}_C$ in the
following.
%
Figures \ref{fig:Feynman1} and \ref{fig:Feynman2} show the
representative Feynman diagrams for the $W_C$ associated production
with a new heavy particle and with a SM particle, respectively.
Between the two mechanisms (associated and pair) for production in Fig.~\ref{fig:Feynman1},
the production rate $bg(\bar{b}g)\rightarrow W^{\pm}_C t^{(1)}$ is clearly
higher than that of $b\bar{b}\rightarrow W^{+}_C W^{-}_C$, due to
its lower kinematical threshold and higher gluon luminosity
in the proton. A similar argument also applies in Fig.~\ref{fig:Feynman2}
in favor of the production $bg\rightarrow  W_C^\pm\ t$.

\begin{figure}
\centerline{\epsfig{file=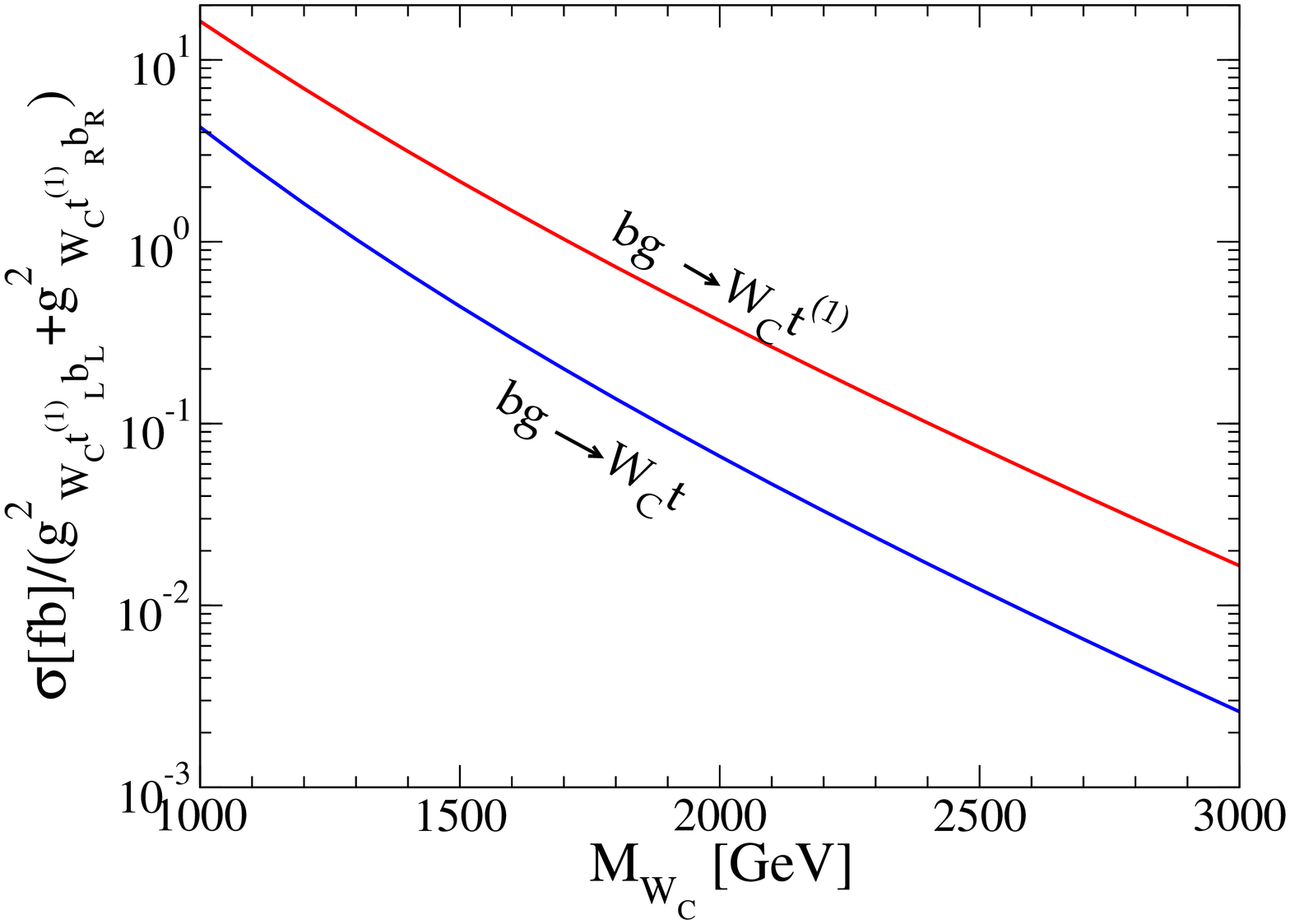,width=6.8cm,angle=-90}
\epsfig{file=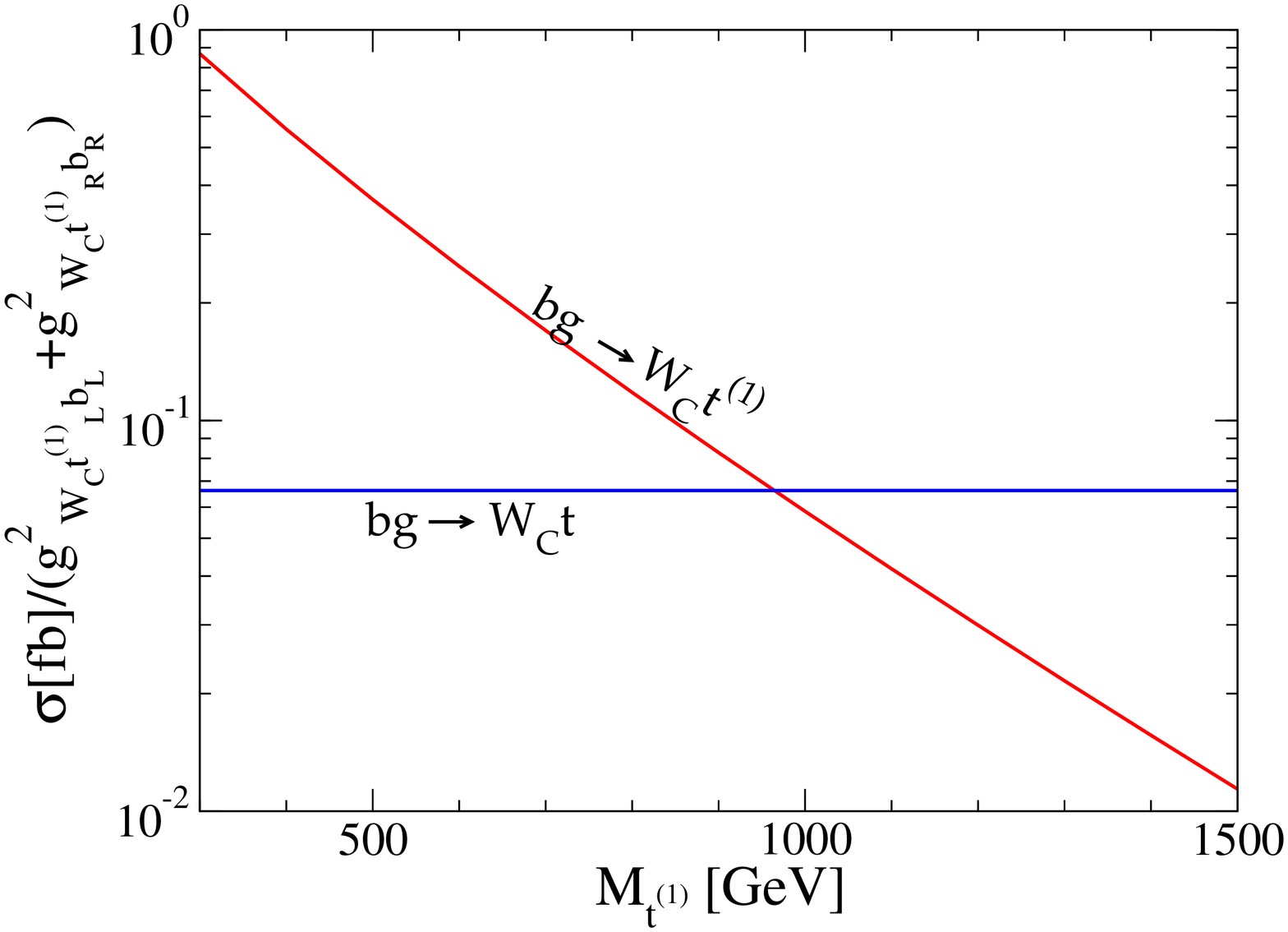,width=6.8cm,angle=-90} } \caption{Cross
section at the LHC (14 TeV) for $pp \rightarrow W^{\pm}_C t^{(1)},\ W^{\pm}_C t$
versus their masses (a) $M_{W_C}$ and (b) $M_{t^{(1)}}$.  The coupling
ratio $g_{W_Ct^{(1)}_Lb_L}/g_{W_Ct_Lb_L}=5$ is fixed. The small
couplings $g_{W_Ct^{(1)}_Rb_R}$ and $g_{W_Ct_Rb_R}$ are turned off,
and order-one coupling square $g^2_{W_Ct^{(1)}_Lb_L}$ is factored
out.} \label{fig:pro}
\end{figure}

In Fig.~\ref{fig:pro}, we show the total cross sections for these
two processes $bg \rightarrow W^{\pm}_C t^{(1)},\ W^{\pm}_C t$, as a
function of their masses (a) $M_{W_C}$ and (b) $M_{t^{(1)}}$. The
t-channel contribution dominates over the s-channel one.
We turn off small couplings $g_{W_Ct^{(1)}_Rb_R}$ and
$g_{W_Ct_Rb_R}$, fix the relative size of couplings as
$g_{W_Ct^{(1)}_Lb_L}/g_{W_Ct_Lb_L}=5$, and factor out the order one
coupling $g_{W_Ct^{(1)}_Lb_L}$.
Comparing these two processes, we see that $W^{\pm}_C t^{(1)}$
production wins due to the stronger coupling as in Fig.~\ref{fig:pro}(a);
while $W^{\pm}_C t$ production wins for the phase space when
$M_{t^(1)} > 1$ TeV, as in Fig.~\ref{fig:pro}(b). The cross sections can
be typically of the order of a fraction of fb in the mass range of our interest.
Since our goal is to explore the reach of LHC on discovering $W_C$, we
will focus on the low mass region of $M_{t^{(1)}}$, where the associated
production $bg(\bar{b}g)\rightarrow W^{\pm}_C t^{(1)}$
dominates among the various non-resonant production channels.

We
estimate, based on appropriate rescaling of numbers in
Fig. 4 (a) of 1st reference in \cite{Agashe:2007ki}
or Fig. 7 of reference \cite{Han:2005ru} for example,
that the
resonant production of
coset gauge bosons via light quarks might be comparable
to the above associated production, but the
former channel suffers (as discussed at the
end of section \ref{sec:two-site})
from a pollution from production of the
KK $W$.
%
%
On the other hand, it is easy to see that a
similar pollution for associated $W_C$ production is negligible:
note that  (as discussed earlier) KK $W$ does
{\em not} couple to $b_L$ (doublet) and light $t^{ (1) }$ (singlet) {\em before} EWSB, i.e., the
pollution from KK $W$
in this channel is suppressed by EWSB.\footnote{The coupling
of KK $W_R$, i.e., the charged
gauge boson of $SU(2)_R$, to $t^{(1)}_L$ and $b_L$ is similarly suppressed compared to that of $W_C$ to $t^{(1)}_L$ and $b_L$, again
since $t^{(1)}$ is mostly singlet of $SU(2)_L\times SU(2)_R$ in the part of parameter space we are considering. Actually, there is another top KK mode living in the bi-doublet representation of $SU(2)_L\times SU(2)_R$, which has a coupling to $W_R$ (but not to KK $W_L$) and $b_L$
which is similar in size to the coupling of coset gauge bosons to
$b_L$ and the singlet $t^{(1)}$. However, the mass of this bi-doublet KK top is $\sim 1.4$ times higher than that of
the singlet $t^{(1)}$ for the point we are considering, leading to the cross section for associated production of $W_R$ (via
exchange of the bi-doublet top KK mode) being suppressed (relative
to that for coset gauge boson with singlet top KK mode exchange), for the case when $W_R$ and
$W_C$ have the same masses. Again, compare this situation to the
pollution encountered in the resonant production of coset gauge bosons mentioned above.}
So, we will consider only associated production of
$W_C$ in this paper.
For the purpose of illustration, we choose
\begin{equation}
\label{eq:wct1mass} M_{W_C}=2~ {\rm TeV},~~ M_{t^{(1)}}=500~ {\rm
GeV}
\end{equation}
as the reference point, and explore the dependence on the masses later.

\subsection{Search of $W_C$ at LHC: Signals and Backgrounds} \label{subsec:signal}

To further quantify the search of $W_C$ at the LHC, we fix to one
point in the parameter space and study the signals and background in
detail. The couplings corresponding to this parameter point is shown
in Table \ref{couplingtable}. We use the $M_{W_C}$ and $M_{t^{(1)}}$
as in Eq.~(\ref{eq:wct1mass}).
%
The cases with other couplings and masses can be estimated
by a proper scaling according to the production cross section shown
in Fig.~\ref{fig:pro}. We use the CTEQ6.1L parton distribution
functions \cite{Pumplin:2002vw}. We concentrate on the dominant
channels of production of $W_C$ and its decay
\begin{equation}
bg \rightarrow W_C~t^{(1)} \to b t^{(1)}\  t^{(1)}.
\end{equation}
We consider all the decay channels
of $t^{(1)}$ as in Eq.~(\ref{eq:t1decchns}), which result in different
signals, as we will study in detail below.

\subsubsection{$t^{(1)} t^{(1)} \rightarrow bW,bW$} \label{subsec:bwbw}

We first consider the case with both $t^{(1)}$'s decaying to $b~W$:
\begin{equation}
bg \rightarrow W_C~t^{(1)} \rightarrow b+2t^{(1)} \rightarrow 3b+2W,
\end{equation}
whose branching fraction is a product of three factors\footnote{We take
the branching fractions of Br$(t^{(1)} \rightarrow b W)$ as $50\%$ in
this estimate. This is a general feature for large $M_{t^{(1)}}$
only. It happens to be approximately true for the parameter point we
use for this detailed study, although it corresponds to a small
$M_{t^{(1)}}$.}
\begin{equation}
Br(W_C \rightarrow b t^{(1)}) \times Br(t^{(1)} \rightarrow b W)^2
\approx 90\% \times (50\%)^2 = 22.5\%. \end{equation}

We consider the semileptonic decays of 2 $W$'s where one $W$ decays
as $W \rightarrow l\nu(l=e,\mu)$ while the other as $W \rightarrow
2j$ ($j$ denotes a jet from a light quark). The branching fraction
for this channel is Br$(WW)\approx2/9\times6/9\times2=8/27$ where
the factor 2 is from exchanging the leptonic and hadronic decaying
$W$'s.

The signal of this channel is therefore $l+5j$ with large missing
transverse momentum carried away by a neutrino. The leading background is
\begin{equation}
pp \rightarrow t \bar{t}+ j \rightarrow l \nu + 5j.
\end{equation}
The other background $pp \rightarrow W^+W^-+3j$ is much smaller, not only
because $W^+W^-$ cross-section is smaller than $t\bar{t}$
cross-section but also because the two more QCD jets in this
background introduce another factor of $\alpha^2_s$ suppression. We
will thus only focus on the background of $t \bar{t}+$ 1 QCD jet.

We adopt the event selection criteria with the basic cuts
\cite{cuts}
\begin{eqnarray}
\label{eq:bascuts}
&& P_{T}(l)>25~{\rm GeV}, ~~~ |\eta(l)| < 2.5, \nonumber \\
&& P_{T}(j) > 20~{\rm GeV}, ~~~ |\eta(j)| < 3, \nonumber \\
&&  \displaystyle{\not \hspace{-0.8ex E}}_T > 25~{\rm GeV}, ~~~
\Delta R_{jj, lj}
> 0.4.
\end{eqnarray}
We smear the lepton and jet energy approximately according to
\begin{equation}
\delta E/E = \frac{a}{\sqrt{E/{\rm GeV}}} \oplus b
\end{equation}
where $a_l=13.4\%$, $b_l=2\%$ and $a_j=75\%$, $b_j=3\%$, and
$\oplus$ denotes a sum in quadrature \cite{Abazov:2007ev}. As shown
in Table \ref{tab:bwbw}, the background is much higher than the
signal if only applying  the basic acceptance cuts. However, the signal
has very unique kinematical features that we will utilize next
to suppress the background and to reconstruct the signal.

\begin{table}[tb]
\caption{The cross sections (in fb) for the signal process $pp
\rightarrow W_C~t^{(1)} \rightarrow l\nu +5j$ and SM background $pp
\rightarrow t \bar{t} +j \rightarrow l\nu +5j $, with the cuts and veto
applied consecutively.  Basic cuts refer to those in Eq.~(\ref{eq:bascuts}).
The ``$M_{3j,l\nu j}$ cuts" refers to the cut condition in Eqs.~(\ref{eq:3jcut})
and (\ref{eq:lnjcut}), and ``$M_{3j}$
veto" refers to the veto condition in Eq.~(\ref{eq:3jmassveto}).}
\begin{center}
\label{tab:bwbw}
\begin{tabular}{|c|c|c|c|c|c|}
\hline  & basic cuts & $P^{\rm highest}_T$ & $E_T^{vis}$
& $M_{3j,l\nu j}$ cuts & $M_{3j}$ veto \\
\hline Signal & 0.045 & 0.040 & 0.037 & 0.025 & 0.025 \\
\hline Background & $2.4 \times 10^4$ & 76 & 8.7 & 1.7 & $< 10^{-4}$ \\
\hline
\end{tabular}
\end{center}
\end{table}

One of striking features of the signal is that the $b$ jet from $W_C$
decay is very energetic due to the heavy mass of $W_C$. Among all
the jets, this $b$ jet has the highest $P_T$ in most cases.
Therefore, one can select the highest jet $P_T$ and impose cut on it
\begin{equation}
\label{eq:ptcut} P^{\rm highest}_T > 500 ~ {\rm GeV}.
\end{equation}
Since the effective c.m. energy is quite large in signal for the
heavy particle production, which are in general higher than those in
background, we further impose cut on the scalar sum of the visible
transverse energies of all the jets and $l$
\begin{equation}
\label{eq:Htcut} E_T^{vis} > 1.5 ~{\rm TeV}.
\end{equation}

With the jet of highest $P_T$ identified as the $b$ jet from $W_C$
decay, there are 4 jets remaining. One can identify which three of
them are from $t^{(1)}$ hadronic decay by selecting 3 jets which
give the invariant mass closest to $t^{(1)}$, and require it to
satisfy
\begin{equation}
\label{eq:3jcut} |M_{3j}-M_{t^{(1)}}|< 50~ {\rm GeV},
\end{equation}
where, as discussed earlier, we have assumed that the mass of $t^{(1)}$ is
known from the early search.
Furthermore, the neutrino momentum can be fully reconstructed using
$W$ mass condition $M^2_W = (p_l + p_{\nu})^2$ with a two-fold
uncertainty \footnote{In doing this, the neutrino $P_T$ is fixed to
balance the $P_T$ of $l$ and jets, which have uncertainties due to
smearing. To accommodate this uncertainty, we allow the $M_W$ to be
as large as 120 GeV. It turns out that, with this range of $M_W$,
there are still cases where there exist no solution for neutrino
momentum, and we lose about 1/3 of events in solving for the neutrino
momentum.}. We select the solution which, in combination with the
momenta of $l$ and the other remaining jet, gives the mass closer to
$M_{t^{(1)}}$, and further require it to satisfy\footnote{Here we
assume $M_{t^{(1)}}$ is known. On the other hand, without assuming
this, one can still fix it by requiring there are 3 jets and
$(l,\nu,j)$ having close heavy masses ($\gg m_t$) and identify this
common mass scale as $M_{t^{(1)}}$.}
\begin{equation}
\label{eq:lnjcut} |M_{l\nu j}-M_{t^{(1)}}|< 100 ~{\rm GeV}.
\end{equation}
With these done, the momenta of two $t^{(1)}$'s are reconstructed.
One still do not know which $t^{(1)}$ is from $W_C$ decay, and
should try both of them, in combination with the jet with highest
$P_T$, to reconstruct the $W_C$ mass. Since one of the two is expected to
be the right one, the collection of the events
should point to  $M_{W_C}$ in the mass distribution.

Although this reconstruction procedure is highly efficient, with a
signal efficiency about $56\%$ and the background rejection of a
factor of $10^{-4}$, as seen from Table \ref{tab:bwbw}, it is still not
sufficient to remove the background.
However, beside identifying the characteristics of the signal, we also notice
the features of the background.  One of the
essential and obvious features of the background is that
there are two top quarks in the event, one of which decays to 3 jets.
One can thus require that there be no combination of 3 jets with
invariant mass within the top mass window
\begin{equation}
\label{eq:3jmassveto} |M_{3j}-m_t| > 50\  {\rm GeV}.
\end{equation}
This veto condition is highly efficient on removing the background.

We show, in Table \ref{tab:bwbw}, the cross sections of signal and
background, with the cuts and veto applied consecutively. Both signal
and background are obtained with parton-level Monte Carlo simulations,
with detector effects accounted for by the geometrical acceptance
and the energy smearing as discussed earlier.
We see that the background is essentially
removed with the veto condition of Eq.~(\ref{eq:3jmassveto}) applied.

In fact, the veto on $M_{3j}$ alone would be sufficient to bring this $t\bar tj$
background below the signal.  Our signal reconstruction scheme is nevertheless
effective to single out the signal from the other potential backgrounds and to
obtain the necessary knowledge about the masses of the heavy particle produced.

\subsubsection{$t^{(1)} t^{(1)} \rightarrow bW,th(tZ)$}
\label{subsec:bwth}

We now consider the case with one $t^{(1)}$ decaying to $b~W$ and
the other decaying to $th$ or $tZ$. Since the signals of the final states are
rather similar for these two cases,  we discuss them together. The signal
we are looking for is
\begin{equation}
bg \rightarrow W_C~t^{(1)} \rightarrow b + 2 t^{(1)} \rightarrow
bbWth(Z) \rightarrow l \nu + 7j,
\end{equation}
whose branching fraction (summing over both $t^{(1)} \rightarrow t~h$
and $t~Z$) is a product of three factors
\begin{equation}
2 \times Br(W_C \rightarrow b t^{(1)}) \times Br(t^{(1)} \rightarrow b W)
\times Br(t^{(1)} \rightarrow th,tZ)
 \approx 2 \times 90\% \times (50\%)^2 = 45\%,
\end{equation}
where the factor of 2 is from exchanging the decay mode of two
$t^{(1)}$'s. We consider semileptonic decay of two $W$'s, which
gives branching fraction 8/27 as discussed earlier.

The dominant SM background for this channel is from the
\begin{equation}
pp \rightarrow t \bar{t}+ 3j \rightarrow l \nu + 7j.
\end{equation}
This can be
considered as adding two more QCD jets to the background considered
in section \ref{subsec:bwbw}. Although we expect this analysis directly analogous
to that in the previous session, one may not be able to effectively calculate this
multiple parton final state. Instead, we will thus simply give an
estimate on how this background is suppressed by various cuts based
on what we learn from the Monte Carlo simulations in section
\ref{subsec:bwbw}. To assess the signal/background ratio for this
channel, we compare both the signal and background with those in the
channel ($t^{(1)} t^{(1)} \rightarrow bW,bW$) in the previous
section.

For signal, this channel has larger branching fraction than
the one considered in section \ref{subsec:bwbw}. However, since
there are two more jets in the final state, there are fewer events
surviving the cuts of $\Delta R_{jj, lj} > 0.4$. As shown in Table
\ref{tab:bwtH}, the signal cross-section after basic cuts in this
channel is very close to that in the channel ($t^{(1)} t^{(1)}
\rightarrow bW,bW$).

The cuts on $P^{\rm highest}_T$ and $E_T^{vis}$ are still applicable
to this channel. Again, the jet with the highest $P_T$ is identified
as the $b$ jet from $W_C$ decay. Among the remaining 6 jets, one can
require that there be at least one pair of jets with invariant mass
close to {\it either} $M_h$, which we assume to be 125 GeV, or $M_Z$
\begin{equation}
\label{eq:2jmasscut} |M_{2j} - M_h| < 15~{\rm GeV}~{\rm or}~|M_{2j}
- M_Z| < 15 ~{\rm GeV}
\end{equation}
The two jets from $h$ or $Z$ decay can be identified in this
way\footnote{If there are more than one pair of jets satisfying
this, one selects the pair whose invariant mass is closer to $M_H$
or $M_Z$.}. The rest of the procedure in reconstructing two $t^{(1)}$
momenta is similar to that in the ($t^{(1)} t^{(1)}
\rightarrow bW,bW$) channel.

Among all the remaining 4 jets, we require there exists at least
one combination of 3 jets and $(l,\nu,j)$, which satisfy
\begin{eqnarray}
\label{eq:3jlnjcut} && |M_{3j}-m_t|<50~ {\rm GeV} ~{\rm and}~
|M_{l\nu j}-M_{t^{(1)}}|<100 ~{\rm GeV} \nonumber \\
~{\rm or}~&& |M_{3j}-M_{t^{(1)}}|<50~ {\rm GeV} ~{\rm and}~
|M_{l\nu j}-m_t|<100~ {\rm GeV}.
\end{eqnarray}
Then, one can combine 2 jets which falls into the $h$ or $Z$ mass
region with the cluster of either 3 jets or $(l,\nu,j)$, whichever
falls into the $M_t$ region, and require this 5 jets or $(l,\nu,3j)$
has mass close to $M_{t^{(1)}}$
\begin{equation}
\label{eq:5jln3jcut} |M_{5j}-M_{t^{(1)}}|<50~ {\rm GeV} ~{\rm or}~
|M_{l\nu 3j}-M_{t^{(1)}}|<100~ {\rm GeV}.
\end{equation}
At this stage, the momenta of two $t^{(1)}$'s are reconstructed,
and one can go further to reconstruct $W_C$ mass by trying both
$t^{(1)}$'s with the jet with the highest $P_T$. Again, the
correlation among events should point to the correct $M_{W_C}$.

Since the top quark also appears in the signal of this channel, the
simple veto on top mass (used in section \ref{subsec:bwbw})
is not applicable here. However, the
background in this channel is sufficiently reduced to be below the
signal with the above cuts applied. This is because there are two
more suppressions in this channel. First, the $t \bar{t}+ 3j$
background is further suppressed by $\alpha^2_s \approx 10^{-2}$
compared to that of $t \bar{t}+ 1j$ background of the ($t^{(1)}
t^{(1)} \rightarrow bW,bW$) channel due to the appearance of two
more QCD jets. Second, the extra cut on $M_{2j}$ in Eq.
(\ref{eq:2jmasscut}) also introduce another factor of suppression,
which we estimate it to be a factor of $10^{-1}$.

The cross-sections of signal and background in this channel, with
cuts applied successively, are summarized in Table \ref{tab:bwtH}.
The cross-section of signal is based on our parton-level
Monte Carlo simulation, and
that of background is obtained from an estimate based on the
cross section of the background in the
$t^{(1)} t^{(1)} \rightarrow bW,bW$ channel in \ref{subsec:bwbw}.

\begin{table}[tb]
\caption{The cross-sections (in fb) for the signal process $pp
\rightarrow W_C~t^{(1)} \rightarrow l\nu +7j$ and SM background $pp
\rightarrow t \bar{t} +3j \rightarrow l\nu +7j $, with the cuts applied
successively. Basic cuts refer to those in Eq.~(\ref{eq:bascuts}).
The cuts on $E_T^{vis}$ and $P^{\rm highest}_T$ refer to Eqs.~(\ref{eq:ptcut})
and (\ref{eq:Htcut}). The ``$M_{2j}$ cut" refer to Eq.~(\ref{eq:2jmasscut}).
The ``$M_{3j,l\nu j}$, $M_{5j,l\nu 3j}$ cuts" refer to Eqs.~(\ref{eq:3jlnjcut})
and (\ref{eq:5jln3jcut}). }
\begin{center}
\label{tab:bwtH}
\begin{tabular}{|c|c|c|c|c|c|}
\hline  & basic cuts & $P^{\rm highest}_T$ & $E_T^{vis}$
& $M_{2j}$ cut & $M_{3j,l\nu j}$, $M_{5j,l\nu 3j}$ cuts   \\
\hline Signal & 0.041 & 0.037 & 0.035 & 0.033 & 0.011 \\
\hline Background & $2.4 \times 10^2$ & 0.76 & 0.087 & 0.0087 & $ < 0.001 $ \\
\hline
\end{tabular}
\end{center}
\end{table}

\subsubsection{$t^{(1)} t^{(1)} \rightarrow t h(t Z),t h(t Z)$}

Finally, we consider the channel with both $t^{(1)}$ decaying to $t
h(t Z)$
\begin{equation}
bg \rightarrow W_C~t^{(1)} \rightarrow b + 2 t^{(1)}  \rightarrow b~
t~ h(Z)~ t~ h(Z) \rightarrow l ~\nu + 9j,
\end{equation}
whose branching fraction (summing over both $t^{(1)} \rightarrow t~h$
and $t~Z$) is a product of three factors
\begin{equation}
Br(W_C \rightarrow b t^{(1)}) \times Br(t^{(1)} \rightarrow th,tZ)^2
 \approx 90\% \times (50\%)^2 = 22.5\%.
\end{equation}
We will again use  semileptonic decay of two $W$'s
whose
branching fraction is 8/27
as discussed earlier.

The dominant SM background for this channel is from the $pp
\rightarrow t \bar{t}+ 5j \rightarrow l \nu + 9j$. We will simply
give an estimate on how this background is suppressed by various
cuts based on what we learned from the study based on Monte Carlo
simulations in section \ref{subsec:bwbw}.

The branching fraction of this channel is similar to the ($t^{(1)}
t^{(1)} \rightarrow bW,bW$) channel. However, with four more jets in
the signal event, the cross-section is drastically reduced after
imposing basic cuts that involve $\Delta R_{jj, lj} > 0.4$.

On the other hand, the background is smaller by a factor of
$\alpha^4_s \approx 10^{-4}$ than in the ($t^{(1)} t^{(1)} \rightarrow
bW,bW$) channel. Some of the cuts discussed previously are still
applicable in this channel. They include the cuts on $E_T^{vis}$, $P^{\rm
highest}_T$, and $M_{2j}$ (we require that there are at least two
pairs of jets that satisfy Eq.~(\ref{eq:2jmasscut}) in this
channel).

Similarly to the other two channels, the two $t^{(1)}$
momenta can be reconstructed and the $W_C$ mass can be obtained from
correlations among events. We skip the details of this procedure here
since it should be clear from discussion in the previous
sections. One should note that the cuts on 3 jets and $(l,\nu,j)$
invariant mass in Eq.~(\ref{eq:3jlnjcut}) should be replaced by
\begin{equation}
\label{eq:3jlnjcutother} |M_{3j}-m_{t}|<50 ~ {\rm GeV}~{\rm and}~
|M_{l\nu j}-m_{t}|<100 ~ {\rm GeV},
\end{equation}
since there are two top quarks.

We show, in Table \ref{tab:tHtH}, the signal and background with
the cuts applied successively. Again, the cross-section of signal is
based on our parton-level Monte Carlo simulation,
and that of background is based on our estimate built upon
section \ref{subsec:bwbw}.

\begin{table}
\caption{The cross-sections (in fb) for the signal process $pp
\rightarrow W_C~t^{(1)} \rightarrow l\nu +9j$ and SM background $pp
\rightarrow t \bar{t} +5j \rightarrow l\nu +9j $, with the cuts applied
successively. Basic cuts refer to those in Eq.~(\ref{eq:bascuts}).
The $P_T$ cuts on $P^{\rm highest}_T$ and $E_T^{vis}$ follow
Eqs.~(\ref{eq:ptcut}) and (\ref{eq:Htcut}), and invariant mass cuts on
$M_{2j}$, $M_{3j,l\nu j}$, and $M_{5j,l\nu 3j}$ follows
Eqs.~(\ref{eq:2jmasscut}), (\ref{eq:3jlnjcutother}) and (\ref{eq:5jln3jcut}). }
\begin{center}
\label{tab:tHtH}
\begin{tabular}{|c|c|c|c|c|c|}
\hline  & basic cuts & $P^{\rm highest}_T$ & $E_T^{vis}$
& two $M_{2j}$ cuts & $M_{3j,l\nu j}$, $M_{5j,l\nu 3j}$ cuts\\
\hline Signal & 0.0081 & 0.0070 & 0.0068 & 0.0064 & 0.0023 \\
\hline Background & $2.4 $ & 0.0076 & 0.00087 & $<8.7 \times 10^{-6}$ & $<1.0 \times 10^{-6}$\\
\hline
\end{tabular}
\end{center}
\end{table}

\subsubsection{Summarizing all channels}

According to the study of each channel presented above,
the background
in each channel can be sufficiently suppressed after imposing
various cut and veto criteria. For reader's convenience, we
reiterate and summarize the cuts condition of all three channels
here again.

Common cuts for all three channels:
\begin{itemize}
\item[(a)] Basic cuts as in Eq.~(\ref{eq:bascuts});
\item[(b)] $P_T$ cuts: $P^{{\rm highest}}_T > 500$ GeV, $E_T^{vis} > 1.5 $ TeV;
\end{itemize}

Specific cuts in each channels:
\begin{itemize}
\item[(i)]Channel I ($l+5j+\displaystyle{\not \hspace{-0.8ex E}}_T$)
\begin{itemize}

\item[(c)] $M_{3j,l\nu j}$ cuts: with the jet with highest $P_T$
excluded, requiring there exists such a combination of 3 jets and
$(l,\nu,j)$ satisfying the invariant mass cut
\begin{equation}
|M_{3j}-M_{t^{(1)}}|< 50 ~{\rm GeV},~|M_{l\nu j}-M_{t^{(1)}}|< 100
~{\rm GeV}
\end{equation}
\item[(d)] $M_{3j}$ veto: requiring there exist no combination of 3
jets that has invariant mass close to top mass
\begin{equation}
|M_{3j}-m_t|> 50 ~{\rm GeV}
\end{equation}
\end{itemize}

\item[(ii)]Channel II ($l+7j+\displaystyle{\not \hspace{-0.8ex E}}_T$)
\begin{itemize}
\item[(c)] $M_{2j}$ cut: with the jet with highest $P_T$
excluded, requiring there exists at least one pair of jets with mass
close to $h$ or $Z$
\begin{equation}
|M_{2j}-M_{h}|< 15 ~{\rm GeV} ~{\rm or}~|M_{2j}-M_{Z}|< 15 ~{\rm
GeV}
\end{equation}
\item[(d)] $M_{3j,l\nu j}$ cuts: within the remaining 4 jets,
requiring there exists at least one combination of 3 jets and
$(l,\nu,j)$ that satisfy
\begin{eqnarray}
\label{eq:3jlnjcut} && |M_{3j}-m_t|<50~ {\rm GeV} ~{\rm and}~
|M_{l\nu j}-M_{t^{(1)}}|<100 ~{\rm GeV} \nonumber \\
~{\rm or}~&& |M_{3j}-M_{t^{(1)}}|<50~ {\rm GeV} ~{\rm and}~
|M_{l\nu j}-m_t|<100~ {\rm GeV}.
\end{eqnarray}
\item[(e)] $M_{5j,l\nu 3j}$ cuts: combining 2 jets that falls into the $h$ or $Z$ mass
region with the cluster of either 3 jets or $(l,\nu,j)$, whichever
falls into the $M_t$ region, and requiring this 5 jets or
$(l,\nu,3j)$ has mass close to $M_{t^{(1)}}$
\begin{equation}
\label{eq:5jln3jcut} |M_{5j}-M_{t^{(1)}}|<50~ {\rm GeV} ~{\rm or}~
|M_{l\nu 3j}-M_{t^{(1)}}|<100~ {\rm GeV}.
\end{equation}
\end{itemize}

\item[(iii)]Channel III ($l+9j+\displaystyle{\not \hspace{-0.8ex E}}_T$)
\begin{itemize}
\item[(c)] $M_{2j}$ cut: with the jet with highest $P_T$
excluded, requiring there exists at least two pairs of jets with
mass close to $h$ or $Z$
\begin{equation}
|M_{2j}-M_{h}|< 15 ~{\rm GeV} ~{\rm or}~|M_{2j}-M_{Z}|< 15 ~{\rm
GeV}
\end{equation}
\item[(d)] $M_{3j,l\nu j}$ cuts: within the remaining 4 jets,
requiring there exists at least one combination of 3 jets and
$(l,\nu,j)$ that satisfy
\begin{eqnarray}
\label{eq:3jlnjcut} && |M_{3j}-m_t|<50~ {\rm GeV} ~{\rm and}~
|M_{l\nu j}-m_t|<100 ~{\rm GeV} .
\end{eqnarray}
\item[(e)] $M_{5j,l\nu 3j}$ cuts: combining 2 pairs of jets which falls into the $h$ or $Z$ mass
region with 3 jets and $(l,\nu,j)$, and requiring that in one of the
two possible combinations, both 5 jets and $(l,\nu,3j)$ have
invariant mass satisfying
\begin{equation}
\label{eq:5jln3jcut} |M_{5j}-M_{t^{(1)}}|<50~ {\rm GeV} ~{\rm and}~
|M_{l\nu 3j}-M_{t^{(1)}}|<100~ {\rm GeV}.
\end{equation}
\end{itemize}

\end{itemize}

\begin{figure}
\centerline{\epsfig{file=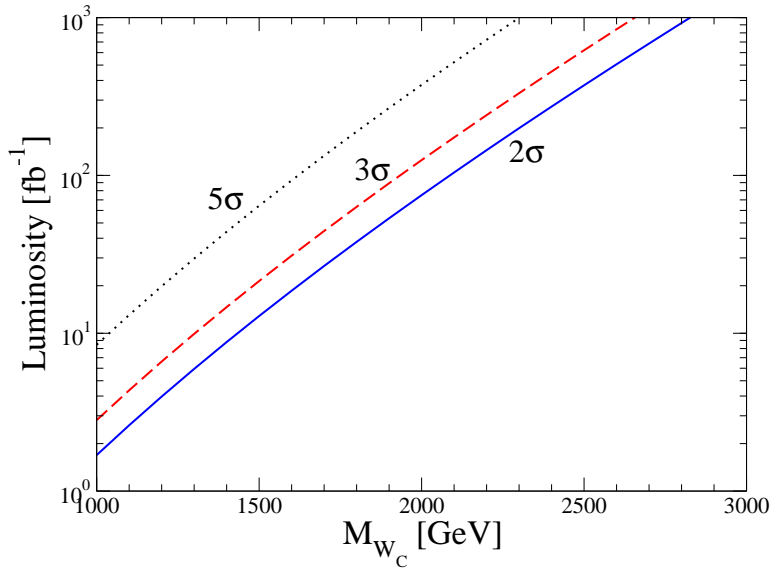,width=9.0cm,angle=-90} } \caption{The
luminosity needed for a 2$\sigma$(blue solid line), 3$\sigma$(red
dashed line), and 5$\sigma$(black dotted line) discovery of $W_C$ at
LHC (14 TeV) as a function of $M_{W_C}$.} \label{fig:lum}
\end{figure}

To assess the discovery potential, we
combine the number of events in
%
%
all the channels
based on the above study of the benchmark point in the parameter space
(with $M_{ W_C } = 2$ TeV).
The luminosity needed at the LHC (14
TeV) for a $95\%$(2$\sigma$), $99.7\%$(3$\sigma$), and
$99.9999\%$(5$\sigma$) c.l. discovery of $W_C$, which implies about
3, 5, and 15 events, respectively (assuming that the background is negligible,
as is the case here),
can then be determined.
For other $M_{ W_C }$ masses, we
rescale this total number of events from the $2$ TeV
case, based on the dependence of cross-section on $M_{ W_C }$ as in
Fig.~\ref{fig:pro}.
These results are
displayed in Fig. \ref{fig:lum}.
Conversely,
for a luminosity of 1000 fb$^{-1}$, the reach in
$M_{W_C}$ is 2.3 TeV at 5$\sigma$, 2.6 TeV at 3$\sigma$, and 2.8 TeV
at 2$\sigma$.


\section{Discussions and Conclusions}
\label{conclude}

During the past decade, the framework of a warped extra dimension
with the SM fields propagating in it has emerged as an attractive
extension of the SM due to its potential for solving both the
Planck-weak and flavor hierarchy problems of the SM.
Moreover, naturalness motivates obtaining SM Higgs as a by-product
of extending the $5D$ EW gauge symmetry beyond the SM one
and breaking it down to the SM near the TeV brane.

In this paper, we first give a full-fledged presentation for the formalism
involving the coset gauge bosons in this framework, {\it i.e.}, the extra (beyond SM-type) EW gauge bosons which are
characteristically doublets under SU(2)$_L$.
We have then performed a study of LHC signals for the coset gauge boson $W_C$.
We have developed a judicial and complex  set of kinematical cuts to optimize the
signal-to-background ratios.
We have found that discovery of these gauge bosons at the LHC  is very
challenging. The primary reason for this is due to their unique gauge quantum numbers,
so that the
coset gauge bosons do not couple at leading order (in Higgs vev) to two SM particles, whether
gauge bosons or fermions. Thus the $s$-channel resonant production
of coset gauge bosons is quite suppressed, making production in
association with KK top an important mechanism (assuming that the KK
top can be as light
$\sim 500$ GeV,
%
%
which does indeed happen quite commonly in this framework) and thus
we focused on this channel in this paper.
This associated production experiences phase space suppression.

On the other hand, the
advantage of this feature of the couplings of coset gauge bosons
is that their two-body decays to SM particles are also suppressed. The
leading decay for charged coset gauge bosons is to $t^{(1)} b$,
and in turn,  $t^{(1)}  \to bW$ or $t h,\ t Z$.
Thus the final states are richer than just two SM particles,
enabling separation from leading SM backgrounds, such as $t \bar{b},\ t\bar t$.
Based on such an analysis, we have estimated the $3\sigma$ reach for
coset gauge bosons to be $\sim 2\ (2.6)$ TeV with $\sim 100\ (1000)$
fb$^{ -1 }$ of luminosity.
We notice, however, that typical models suggest that the mass
scale of these gauge bosons is at least $\sim 3$ TeV. This expectation
is based on an {\em indirect}
limit from precision EW observables (direct bound on coset gauge boson mass
being weaker), namely, due to the
masses of coset gauge boson and those of the KK excitations of SM gauge bosons
being related and the latter being directly constrained by precision EW observables.

The same feature of the coset gauge boson couplings also makes their
signals distinct from those of other heavy EW gauge bosons as follows.
Consider the signals for the EW KK gauge bosons with the same
quantum numbers as the SM gauge bosons within the same framework,
{\it i.e.}, KK $Z,\ W$ and $\gamma$ \cite{Agashe:2007ki}. They
couple at leading order to two SM particles, for example, to third
generation quarks, $WW/WZ$\footnote{Note that there are actually
more than one neutral and charges states so that one mass eigenstate
(i.e., admixture of gauge eigenstates) might have suppressed
couplings due to cancelation between its various components, but
then the other (almost degenerate) state does not have such
suppressed couplings.} and even to two light quarks (but typically
with suppressed couplings compared to the SM ones), unlike coset
gauge bosons which do not couple to $WW/WZ$ or to two SM quarks at leading order
in Higgs vev.
Thus, the production cross-section is larger for
KK $W/Z/\gamma$ than for coset gauge bosons of the same mass, but the decay
channels of KK $W/Z$ are not as rich as for coset gauge bosons.
%
%
%
We can also compare to signals for Little Higgs models,
where $Z^{ \prime }$/$W^{ \prime }$, without $T$-parity,
generically do couple to (and hence can be produced by or decay into)
$WW/WZ$ or to two SM light fermions \cite{Han:2003wu}, resulting in distinct
signatures from those of the coset gauge bosons.
Similarly, $4D$ Left-Right symmetric models have a $W^+_R$ which does {\em not}
couple to $W/Z$ at leading order but it does couple to
two SM right-handed quarks and hence it can be produced
by light quark-antiquark annihilation \cite{Keung:1983uu}
and can decay into $t_R \bar{b}_R$.
Finally, we must note that for many of the gauge sector extensions to
include $Z^{ \prime }/W^{ \prime }$
(KK $W/Z$, photon in
warped extra dimensional framework being
notable exceptions),
their leptonic decays are always the gold-plated signatures,
which is absent for the coset gauge boson searches.

We envisage the following sequence of events if this
framework were realized in Nature:  It is likely that a light KK top $t^{(1)}$
will be the first new particles to be discovered, with
possibly less than
%
%
10 fb$^{-1}$ luminosity.
With about $100$ fb$^{-1}$ of integrated
luminosity, it is the turn of the KK gluon next via its decay to $t \bar{t}$,
followed closely by the KK $W/Z$ via both gauge boson $WW/WZ$ and fermion $t \bar{b}/t \bar{t}$ final
states\footnote{although KK gluon decays to $t \bar{t}$ could be a
``background'' for KK $Z$ in the $t \bar{t}$ channel.}. Finally with even higher luminosity, the
coset gauge bosons can be searched for using final states with
top/bottom/$W$ (i.e., like some decays of KK $W/Z$, but with extra
particles), but with no corresponding signal/excess in $WW/WZ$ final states.
%
%
Although the signatures of the new particles
in the warped extra dimensional framework (including those of coset gauge bosons
in the more natural versions)
are qualitatively distinctive, it is clear that their detailed analyses at the LHC
would be
required in order to
establish this attractive theoretical framework.

\section*{Acknowledgments}

We would like to thank Roberto Contino and Raman Sundrum for
valuable discussions. K.A.~was supported in part by an NSF grant
No.~PHY-0652363. T.H.~was supported by the US DOE under contract
DE-FG02-95ER40896. Y.L.~was supported by the US DOE under contract
DE-FG02-08ER41531.

\appendix
\section{$SO(5)$ generators and group algebras}\label{SO5}

The commutation relations of $SO(5)$ generators are given by
\begin{eqnarray}
[T_L^a, T_L^b] = i \epsilon^{abc}T_L^c, \quad [T_R^a, T_R^b] = i
\epsilon^{abc}T_R^c, \quad [T_L^a, T_R^b] = 0 \\ \nonumber
[T^{\hat{a}}, T^{\hat{b}}] = \frac{i}{2}\epsilon^{abc}(T_L^c  +
T_R^c), \quad [T^{\hat{a}}, T^{\hat{4}}] = \frac{i}{2}(T_L^a - T_R^a)\\
\nonumber [T_{L,R}^a, T^{\hat{b}}] =
\frac{i}{2}(\epsilon^{abc}T^{\hat{c}} \pm \delta^{ab}T^{\hat{4}}),
\quad [T_{L,R}^a, T^{\hat{4}}] = \mp\frac{i}{2}T^{\hat{a}}
\end{eqnarray}
For $\mathbf{5}$ representation, the generators are
\begin{eqnarray}
T_{L,R}^1 &=& \frac{-i}{2}\left(\begin{array}{ccccc} 0 & 0 & 0 & \pm 1 & 0\\
0 & 0 & 1 & 0 & 0\\ 0 & -1 & 0 & 0 & 0 \\ \mp 1 & 0 & 0 & 0 & 0 \\ 0
& 0 & 0 & 0 & 0
\end{array}\right), \quad T_{L,R}^2 = \frac{-i}{2}\left(\begin{array}{ccccc} 0 & 0 & -1 & 0 & 0\\
0 & 0 & 0 & \pm 1 & 0\\ 1 & 0 & 0 & 0 & 0 \\ 0& \mp 1 & 0  & 0 & 0 \\
0 & 0 & 0 & 0 & 0
\end{array}\right)\\ \nonumber
T_{L,R}^3 &=& \frac{-i}{2}\left(\begin{array}{ccccc} 0 & 1 & 0 & 0 & 0\\
-1 & 0 & 0 & 0 & 0\\ 0 & 0 & 0 & \pm 1 & 0 \\ 0 & 0 & \mp 1 & 0 & 0 \\
0 & 0 & 0 & 0 & 0
\end{array}\right), \quad T^{\hat{1}} = \frac{-i}{\sqrt{2}}\left(\begin{array}{ccccc} 0 & 0 & 0 & 0 & 1\\
0 & 0 & 0 & 0 & 0\\ 0 & 0 & 0 & 0 & 0 \\ 0 & 0 & 0 & 0 & 0 \\ -1 & 0
& 0 & 0 & 0
\end{array}\right)\\ \nonumber
T^{\hat{2}} &=& \frac{-i}{\sqrt{2}}\left(\begin{array}{ccccc} 0 & 0 & 0 & 0 & 0\\
0 & 0 & 0 & 0 & 1\\ 0 & 0 & 0 & 0 & 0 \\ 0 & 0 & 0 & 0 & 0 \\ 0 &
-1& 0 & 0 & 0
\end{array}\right),\quad T^{\hat{3}} = \frac{-i}{\sqrt{2}}\left(\begin{array}{ccccc} 0 & 0 & 0 & 0 & 0\\
0 & 0 & 0 & 0 & 0\\ 0 & 0 & 0 & 0 & 1 \\ 0 & 0 & 0 & 0 & 0 \\ 0 & 0
& -1 & 0 & 0
\end{array}\right)\\ \nonumber
T^{\hat{4}} &=& \frac{-i}{\sqrt{2}}\left(\begin{array}{ccccc} 0 & 0 & 0 & 0 & 0\\
0 & 0 & 0 & 0 & 0\\ 0 & 0 & 0 & 0 & 0 \\ 0 & 0 & 0 & 0 & 1 \\ 0 & 0
& 0 &-1 & 0
\end{array}\right)
\end{eqnarray}

\section{KK Decomposition and Spectral Functions of Gauge Bosons and
Fermions}\label{KK decomp}

In this appendix, we review the KK decomposition and spectral
functions of gauge bosons and fermions in the
model\cite{Medina:2007hz}. First, we define base functions for gauge
bosons
\begin{eqnarray}
C_A(z, m_n) = \frac{\pi m_n}{2} z\left[J_1(m_n z) Y_0(m_n) - J_0(m_n
) Y_1(m_n z)\right]\\
S_A(z, m_n) = \frac{\pi m_n}{2} z\left[J_1(m_n ) Y_1(m_n z) -
J_1(m_n z) Y_1(m_n)\right]
\end{eqnarray}
These base functions are like cosine and sine in flat extra
dimension. $C_A(z, m_n)$ has ``$+$'' boundary condition on the UV
brane and $S_A(z, m_n) $ has ``$-$'' boundary condition on the UV
brane. They both solve the equations of motion for gauge boson
wavefunctions with vanishing Higgs vev. Then the wavefunctions of
gauge fields with vanishing Higgs vev are easy to write down in
terms of these base functions
\begin{eqnarray}
f^{a_L}_n (z, 0) = C_{a_L,n} C_A(z,m_n)\\
f^{\hat{a}}_n (z, 0) = C_{\hat{a},n} S_A(z,m_n)\\
f^{Y}_n (z, 0) = C_{Y,n} C_A(z,m_n)\\
f^{a_R}_n(z,0) = C_{a_R, n} S_A(z,m_n)
\end{eqnarray}
The wavefunctions in the presence of background $\langle
A_z^{\hat{4}}\rangle$ can be obtained by using Eq.~(\ref{Eq trans}).
For simplicity, we define
\begin{equation}
\theta_G(z) =  \frac{v(z^2-R^2)}{\sqrt{2}f_h(z_\pi^2-R'^2)}
\end{equation}
Then $\Omega(z,v) = e^{-i\sqrt{2} \theta_G(z) T^{\hat{4}}}$. Use the
representation of $SO(5)$ generators in Appendix \ref{SO5}, we
obtain
\begin{eqnarray}\label{Eq gauge functions with nonvanishing vev}
f^{1_L}(v) &=& \frac{1}{2}(1+ \cos{\theta_G}) C_{1_L} C_A(z) +
\frac{1}{2}(1- \cos\theta_G) C_{1_R}S_A(z) + \frac{\sqrt{2}}{2} \sin
\theta_G C_{\hat{1}}S_A(z)\\
f^{2_L}(v) &=& \frac{1}{2}(1+ \cos{\theta_G}) C_{2_L} C_A(z) +
\frac{1}{2}(1- \cos\theta_G) C_{2_R}S_A(z) + \frac{\sqrt{2}}{2} \sin
\theta_G C_{\hat{2}}S_A(z)\\
f^{3_L}(v) &=& \frac{1}{2}(1+ \cos{\theta_G}) C_{3_L} C_A(z) +
\frac{1}{2}(1- \cos\theta_G) [c_\phi C_{3_R} S_A(z) + s_\phi C_Y
C_A(z)]
\\ \nonumber &+& \frac{\sqrt{2}}{2} \sin
\theta_G C_{\hat{3}}S_A(z)\\
f^{1_R}(v) &=& \frac{1}{2}(1- \cos{\theta_G}) C_{1_L} C_A(z) +
\frac{1}{2}(1+ \cos\theta_G) C_{1_R}S_A(z) - \frac{\sqrt{2}}{2} \sin
\theta_G C_{\hat{1}}S_A(z)\\
f^{2_R}(v) &=& \frac{1}{2}(1- \cos{\theta_G}) C_{2_L} C_A(z) +
\frac{1}{2}(1+ \cos\theta_G) C_{2_R}S_A(z) - \frac{\sqrt{2}}{2} \sin
\theta_G C_{\hat{2}}S_A(z)\\
f^{3_R}(v) &=& \frac{1}{2}(1- \cos{\theta_G}) C_{3_L} C_A(z) +
\frac{1}{2}(1+ \cos\theta_G) \left[c_\phi C_{3_R} S_A(z) + s_\phi
C_Y C_A(z)\right]\\ \nonumber &-& \frac{\sqrt{2}}{2} \sin
\theta_G C_{\hat{3}}S_A(z)\\
f^{\hat{1}}(v) &=& \cos \theta_G C_{\hat{1}} S_A(z) + \sin \theta_G
\frac{1}{\sqrt{2}}\left[ C_{1_R} S_A(z) - C_{1_L} C_A(z) \right]\\
f^{\hat{2}}(v) &=& \cos \theta_G C_{\hat{2}} S_A(z) + \sin \theta_G
\frac{1}{\sqrt{2}}\left[ C_{2_R} S_A(z) - C_{2_L} C_A(z) \right]\\
f^{\hat{3}}(v) &=& \cos \theta_G C_{\hat{3}} S_A(z) + \sin \theta_G
\frac{1}{\sqrt{2}}\left[ c_\phi C_{3_R}  S_A(z)+ s_\phi C_Y C_A(z) - C_{3_L} C_A(z) \right]\\
f^{\hat{4}}(v) &=& C_{\hat{4}} S_A(z)\\
f^{X}(v) &=& c_\phi C_Y C_A(z) - s_\phi C_{3_R} S_A(z)
\end{eqnarray}
where the dependence on $z$ and KK number $n$ is not shown
explicitly. The boundary conditions of $f^{a,\hat{a}}(v,z)$ at
$z=R'$ set the eigenvalues $m_n$. We can separate the gauge bosons
in three sectors: (i) $a = 1_L, 1_R, \hat{1}, 2_L, 2_R, \hat{2}$,
these gauge bosons correspond to $W^\pm$ and their KK modes and
coset $W_C$ gauge boson. (ii) $a = 3_L, 3_R, \hat{3}, X$, these
correspond to neutral gauge bosons ($Z$, $\gamma$ and coset $Z_C$
gauge boson). (iii) $a = \hat{4}$, corresponds to the gauge boson
partner of physical Higgs boson. We now study these sectors.
\begin{itemize}
\item (i) $W^{\pm}$ sector.
The boundary conditions on the IR brane are
\begin{eqnarray}
\partial_z f_{i_L}(z_\pi,v) =0\\
\partial_z f_{i_R}(z_\pi,v) = 0\\
f_{\hat{i}}(z_\pi,v) =0
\end{eqnarray}
where $i=1, 2$. These boundary conditions give us
\begin{eqnarray}
&&C_{i_L}\left[ C_A(R') - \sin \theta_G C_A(z_\pi) \theta'_G +
\cos\theta_G C'_A(R')\right] \\ \nonumber &+& C_{i_R}\left[ S'_A(R')
+ \sin
\theta_G S_A(R') \theta'_G - \cos\theta_G S'_A(R')\right]\\
\nonumber &+& C_{\hat{i}}\left[ \sqrt{2} \cos\theta_G \theta'_G
S_A(R') + \sqrt{2} \sin\theta_G S'_A(R') \right]=0\\
\\ \nonumber
&&C_{i_L}\left[ C'_A(R') + \sin \theta_G C_A(R') \theta'_G -
\cos\theta_G C'_A(R')\right] \\ \nonumber &+& C_{i_R}\left[ S'_A(R')
- \sin
\theta_G S_A(R') \theta'_G + \cos\theta_G S'_A(R')\right]\\
\nonumber &-& C_{\hat{i}}\left[ \sqrt{2} \cos\theta_G \theta'_G
S_A(R') + \sqrt{2} \sin\theta_G S'_A(R') \right]=0\\ \\ \nonumber
&&\cos \theta_G C_{\hat{i}} S_A(R') + \sin \theta_G
\frac{1}{\sqrt{2}}\left[ C_{i_R} S_A(R') - C_{i_L} C_A(R') \right]=0
\end{eqnarray}
where all functions are evaluated at $z=R'$. These are linear
algebraic equations for the coefficients $C_{i_L, i_R, \hat{i}}$. To
have a solution on the coefficients $C_{i_L , i_R , \hat{i}}$, we
need to require the determinant to be zero. It gives us
\begin{eqnarray}
 C_A(R') S_A'(R') \sin^2 \theta_G + (2 - \sin^2 \theta_G) C'_A(R')
 S_A(R') = 0
\end{eqnarray}
which can be further simplified to
\begin{eqnarray}
1 + F_W(m_n^2) \sin^2\left( \frac{v}{\sqrt{2}f_h}\right)=0\\
\nonumber F_W(m^2) \equiv \frac{m R'}{2 C'(R', m) S(R', m)}
\end{eqnarray}
Here we defined the form factor of W bosons $F_W(m^2)$. Now we can
see that the spectral function of $W$ boson is
\begin{equation}\label{W spec}
\rho_W(m) = 1 + F_W(m^2)\sin^2\left(\frac{v}{\sqrt{2}f_h}\right)
\end{equation}

\item (ii) $Z, \gamma$ sector. The boundary conditions on the IR
brane are
\begin{eqnarray}
\partial_z f_{3_L}(z_\pi,v) = 0\\
\partial_z[c_\phi f_{3_R}(z_\pi,v) - s_\phi f_X(z_\pi,v)] = 0 \\
\partial_z [ s_\phi f_{3_R}(z_\pi,v) + c_\phi f_{X}(z_\pi,v)  ] = 0\\
f_{\hat{3}}(z_\pi,v) = 0
\end{eqnarray}
These boundary conditions give us
\begin{eqnarray}
&&C_{3_L}\left[ C_A'(R') -\sin\theta_G C_A(R') \theta'_G +
\cos\theta_G C'_A(R')\right]\\ \nonumber &+& C_{3_R}c_\phi\left[
S'_A(R') + \sin \theta_G S(R') \theta'_G - \cos\theta_G S'_A(R')
\right]\\ \nonumber &+& C_Y s_\phi\left[ C'_A(R') + \sin\theta_G
C_A(R') \theta'_G - \cos\theta_G C'_A(R') \right]\\ \nonumber &+&
\sqrt{2}C_{\hat{3}} \left[ \sin\theta_G S'_A(R') + \cos\theta_G
S_A(R') \theta'_G \right]=0\\ \nonumber \\&& C_{3_L}c_\phi[
C'_A(R')-\cos\theta_G C'_A(R') + \sin\theta_G C_A(R') \theta_G']
 \\
\nonumber &+& C_{3_R}[(1+ s_\phi^2) S'_A(R') + c_\phi^2 \cos\theta_G
S'_A(R') - c_\phi^2 \sin\theta_G S_A(R') \theta'_G ] \\
\nonumber &+& C_Y s_\phi c_\phi[-C'_A(R') -\sin\theta_G C_A(R')
\theta'_G + \cos\theta_G C'_A(R')]
 \\
\nonumber &-& C_{\hat{3}}\sqrt{2}c_\phi[ \sin\theta_G S'_A(R') +
\cos\theta_G S_A(R') \theta'_G] = 0\\
\nonumber \\
&&C_{3_L}s_\phi[C'_A(R') -\cos\theta_G C'_A(R') + \sin\theta_G
C_A(R') \theta'_G]\\ \nonumber &+& C_{3_R}s_\phi c_\phi[-S'_A(R')
+ \cos\theta_G S'_A(R') -\sin\theta_G S_A(R') \theta'_G]\\
\nonumber &+& C_Y[(1+c_\phi^2) C'_A(R') + s_\phi^2 \cos\theta_G
C'_A(R') -s_\phi^2 \sin\theta_G C_A(R')\theta'_G]\\ \nonumber &-&
C_{\hat{3}}\sqrt{2}s_\phi[\sin\theta_G S'_A(R') + \cos\theta_G
S_A(R') \theta'_G] = 0\\ \nonumber \\
&&C_{3_L}[-\sin\theta_G C_A(R')] + C_{3_R}[c_\phi \sin\theta_G
S_A(R')]\\ \nonumber &+& C_{Y}[s_\phi \sin\theta_G C_A(R')] +
C_{\hat{3}} [\sqrt{2}\cos\theta_G S_A(R')] = 0
\end{eqnarray}
By requiring the determinant is zero, we get
\begin{eqnarray}
C'_A(R') S'_A(R')\Big\{2 C'_A(R') S_A(R') + \sin^2\theta_G
(1+s_\phi^2)[C'_A(R') S_A(R')- C_A(R')S'_A(R')]\Big\}=0
\end{eqnarray}
$C'_A(R')=0$ gives the spectrum of KK photon and $S'_A(R') = 0$
gives the spectrum of KK $W_{3_R}$. Note that their spectrum does
not depend on the Higgs vev thus does not contribute to the CW
potential. With some simplification we can get the spectral function
for $Z$ boson
\begin{equation}\label{Z spec}
\rho_Z(m) = 1 + F_Z(m^2)\sin^2\left(\frac{v}{\sqrt{2}f_h}\right)
\end{equation}
with the $Z$ boson form factor
\begin{eqnarray}
F_Z(m^2) = \frac{(1 + s_\phi^2) m R'}{2 C'(R', m) S(R', m)}
\end{eqnarray}
\item (iii) $A^{\hat{4}}$ sector. The gauge transformation
$\Omega(z,v)$ does not change the wavefunction of $A^{\hat{4}}$.
Therefore
\begin{equation}
f_n^{\hat{4}}(z,v) = f_n^{\hat{4}}(z,0) = C_{\hat{4},n}S_A(z,m_n)
\end{equation}
Its spectrum is determined by $S_A(R',m_n) = 0$. Since the spectral
function does not depend on the Higgs vev, it will not contribute to
the Higgs potential.
\end{itemize}

For the fermionic section, we define the following base function
\begin{eqnarray}
\tilde{S}^F_{M}(z,m_n) &=& \frac{\pi m_n}{2}
z^{\alpha}[J_\alpha(m_n)Y_{\alpha}(m_n z) -
Y_{\alpha}(m_n)J_{\alpha}(m_n z)]\\
S^F_{\pm M} &=& z^{2\pm M} \tilde{S}^F_{\pm M}\\
\dot{S}^F_{\pm M} &=& \mp \frac{z^{2\pm M}}{m_n} \partial_z
\tilde{S}^F_{\pm M}
\end{eqnarray}
with $\alpha = 1/2 + c$ and $M = -c$. $S_{\pm M}$ and $\dot{S}_{\pm
M}$ satisfy Dirichlet and Neunman boundary conditions respectively
at the UV brane. We can do the following KK decomposition for
fermionic wavefunctions with vanishing Higgs vev
\begin{eqnarray}
F^\Psi_{1L}(z,0) = \left( \begin{array}{c} C_1 S^F_{M_1}\\ C_2
S^F_{M_1}\\C_3 \dot{S}^F_{-M_1}\\ C_4 \dot{S}^F_{-M_1}\\ C_5
S^F_{M_1}
\end{array} \right) ,\quad
F^\Psi_{2R}(z,0) = \left(\begin{array}{c} C_6
S^F_{-M_2}\\ C_7 S^F_{-M_2}\\C_8 {S}^F_{-M_2}\\ C_9 \dot{S}^F_{-M_2}\\
C_{10} \dot{S}^F_{M_2}
\end{array}  \right), \quad
F^\Psi_{3R}(z,0) =  \left( \begin{array}{c} C_{11}
S^F_{-M_3}\\ C_{12} S^F_{-M_3}\\C_{13} {S}^F_{-M_3}\\ C_{14} {S}^F_{-M_3}\\
C_{15} {S}^F_{-M_3}\\ C_{16} S^F_{-M_3}\\ C_{17} S^F_{-M_3}\\
C_{18} S^F_{-M_3}\\  C_{19} S^F_{-M_3}\\ C_{20} \dot{S}^F_{M_3}
\end{array}\right)
\end{eqnarray}
As before, the wavefunctions with non-vanishing Higgs vev is given
by doing gauge transformation Eq.~(\ref{fermion gauge
transformation}).  The boundary terms in Eq.~(\ref{fermion boundary
terms}) give twisted boundary conditions for fermions at the IR
brane
\begin{eqnarray}
\chi_{1R} + M_{B_2}\chi_{3R} &=& 0, \qquad \tilde{t}_{1R} +
M_{B_2}\tilde{t}_{3R} = 0 , \qquad t_{1R} + M_{B_2}t_{3R} = 0\\
\nonumber b_{1R} + M_{B_2}b_{3R} &=& 0, \qquad \hat{t}_{1R} +
M_{B_1}\hat{t}_{2R} = 0, \qquad \chi_{3L} -M_{B_2}\chi_{1L} = 0\\
\nonumber \tilde{t}_{3L} - M_{B_2}\tilde{t}_{1L} &=& 0,\qquad t_{3L}
- M_{B_2}t_{1L} = 0,\qquad b_{3L} - M_{B_2}b_{1L} = 0,\qquad
\hat{t}_{2L} - M_{B_1}\hat{t}_{1L}=0
\end{eqnarray}
The rest of the boundary conditions are not changed
\begin{eqnarray}
(\chi_{2L}, \hat{t}_{2L}, t_{2L}, b_{2L}) = 0\\
(\Xi'_{3L}, T'_{3L}, B'_{3L}, \Xi_{3L}, T_{3L}, B_{3L}) = 0
\end{eqnarray}
This boundary conditions set the mass spectra for fermions. The
calculation for fermionic spectral function is similar to the case
of gauge boson. We do not carry out the calculation here but present
the fermionic form factors and spectral functions here for reference
(for more detailed calculation, see \cite{Medina:2007hz}). The
fermionic form factors are
\begin{eqnarray}
F_b(m^2) &=& - \frac{M_{B_2}^2 S^{F\prime}_{-c_1}}{2
S^F_{c_3}(M_{B_2}^2 S^F_{-c_3}
S^{F\prime}_{-c_1} + S^F_{-c_1} S^{F\prime}_{-c_3})}\\
F_{t_1}(m^2) &=& \frac{F_1(m^2)}{F_0(m^2)}\\
F_{t_2}(m^2) &=& \frac{F_2(m^2)}{F_0(m^2)}\\
F_1(m^2) &=& kz\Big\{M_{B_2}^2 S^F_{c_2} S^F_{-c_3}
S^{F\prime}_{-c_2} +
M_{B_1}^2[2 M_{B_2}^2 S^F_{c_1} S^F_{-c_3} S^{F\prime}_{-c_1}\\
\nonumber &+& S^{F\prime}_{-c_3} +
2S^F_{c_1}S^F_{-c_1}S^{F\prime}_{-c_3} - S^F_{c_2} S^F_{-c_2}
S^{F\prime}_{-c_3}]\Big\}\\
F_2(m^2) &=& - (kz) M_{B_1}^2 S^{F\prime}_{-c_3}\\
F_0(m^2) &=& 2\Big\{M_{B_1}^2 S^F_{c_1}(-1 +
S^F_{c_2}S^F_{-c_2})(M_{B_2}^2 S^F_{-c_3} S^{F\prime}_{-c_1} kz +
S^F_{-c_1} S^{F\prime}_{-c_3} kz) \\ \nonumber &+& S^F_{c_2}
S^{F\prime}_{-c_2}kz\left[M_{B_2}^2(-1 +
S^F_{c_1}S^F_{-c_1})S^F_{-c_3}- \frac{1}{m^2}
S^F_{-c_1}S^{F\prime}_{c_1}S^{F\prime}_{-c_3}\right] \Big\}
\end{eqnarray}
The fermionic spectral functions are given by
\begin{eqnarray} \label{bottom spec}
\rho_b(m) &=& 1 + F_b(m^2) \sin^2\left(\frac{v}{\sqrt{2}f_h}\right),
\\\rho_t(m) &=& 1 + F_{t_1}(m^2)
\sin^2\left(\frac{v}{\sqrt{2}f_h}\right)+ F_{t_2}(m^2)
\sin^4\left(\frac{v}{\sqrt{2}f_h}\right).\label{top spec}
\end{eqnarray}

\section{Suppression of $Z_c b_L \bar{b}_L$
coupling}\label{sec:zbb}

\begin{figure}
\label{zbb}
\includegraphics[scale=0.9]{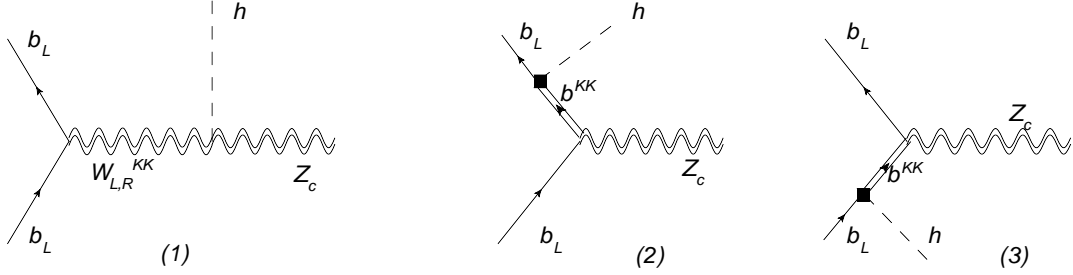}
\caption{Diagrams contributing to the $Z_c \bar{b}b$ at
$\frac{v}{f_h}$ order}\label{fig:zbb}
\end{figure}

In the main text we commented that the couplings of the SM $b$ quark
to the coset $Z_C$ gauge boson are strongly suppressed compared to
the their naive estimates. In this appendix we will explain the
origin of this suppression. From isospin quantum numbers of the
coset gauge bosons
 $(T^3_L,T^3_R)=(\pm \frac{1}{2},\pm\frac{1}{2})$, we see that
 to get  coupling between coset gauge boson and SM fermion we need odd number of Higgs vev insertions.
  In this section we will study only the effects  coming from one Higgs insertion because the
  diagrams with three Higgs insertions will be suppressed due to the additional powers of the $\theta_H^2\equiv\left( \frac{v}{\sqrt2 f_h}\right)^2$.
 The dominant contribution to the  $Z_C b_L \bar{b}_L$ is shown on the Fig.
 \ref{fig:zbb}.
If we only consider individual contributions coming from these
mixings, then the naive estimate of the $Z_C b_L \bar{b}_L$ coupling
will be (we ignore the difference between $\theta_H$ and
$\sin\theta_H$)
\begin{eqnarray}
\sim g_*\left(\frac{1}{2}-c_1\right)\theta_H
\end{eqnarray}
However, to get a more precise estimate, first let us look at the
diagram (1) of Fig.~\ref{fig:zbb}. In this case intermediate gauge
boson can be either $W_{L}^{3,\text{KK}}$ or $W_{R}^{3,\text{KK}}$
(they are the heavy gauge bosons of the generators $T_L^3$ and
$T_R^3$ in two-site language). To analyze the coupling $h Z_{C,\mu}
W^{3, \text{KK}, \mu}_{L,R}$ we will use two-site approach, one can
see that this coupling arises from the covariant derivative of the
$\phi$ field $|D_\mu \phi|^2$ (see Eq.~(\ref{Eq: fphi})). Performing
commutation relations one can show that the couplings of $Z_C$ to
$W_{L}^{3,\text{KK}}$ and $W_R^{3,\text{KK}}$ have opposite signs,
so the effective coupling of $Z_C$ to the SM fermions coming from
diagram(1) of Fig.~\ref{fig:zbb} is proportional to $(T_L^3 -
T_R^3)$ of the $b_L$ field.  But we know that to overcome the
constraint from the shift in $Z\bar{b}_L b_L$ coupling, $b_L$ should
be in such representation of $SU(2)_L\times SU(2)_R$ to have
$T_L^3=T_R^3$ \cite{Agashe:2006at}. This means that in realistic
models of PGB Higgs, contribution to $Z_c b_L \bar{b}_L$ coupling
from diagram (1) of Fig.~\ref{fig:zbb} is zero.

 Now let us look on the diagrams (2) and (3) of Fig.\ref{fig:zbb}.
 Note that $h$ and $Z_C$ are accompanied by the generators $T_c^4$
 and $T_c^3$ of $SO(5)$ respectively. Therefore in this case coupling of the $Z_C$ to the SM $b_L$ is
  proportional to
 \begin{eqnarray}
\bar b_L \left\{T_c^3, T_c^4 \right\}b_L.
 \end{eqnarray}
But one can see that in our model SM $b_L$ lives mostly  in  the
${\bf 5}$  of the SO(5),  then one can check by direct calculation
that
\begin{eqnarray}
\xi_b^\dagger \cdot A^\dagger \cdot\left\{T_c^3, T_c^4 \right\}\cdot
A\cdot\xi_b=0
\end{eqnarray}
where $\xi_b^T=(0,0,0,\psi_b,0)$. This concludes our analysis of the
suppression of the $Z_C b_L \bar{b}_L$ couplings.




\begin{thebibliography}{99}


\bibitem{Randall:1999ee}
  L.~Randall and R.~Sundrum,
  Phys.\ Rev.\ Lett.\  {\bf 83}, 3370 (1999)
  [arXiv:hep-ph/9905221].


\bibitem{Gherghetta:2000qt}
T.~Gherghetta
and A.~Pomarol,
Nucl.\ Phys.\ B {\bf 586}, 141 (2000)
[arXiv:hep-ph/0003129].

\bibitem{review}
  H.~Davoudiasl, S.~Gopalakrishna, E.~Ponton and J.~Santiago,
  arXiv:0908.1968 [hep-ph].


\bibitem{Agashe:2003zs}
K.~Agashe, A.~Delgado, M.~J.~May and R.~Sundrum,
JHEP {\bf 0308}, 050 (2003) [arXiv:hep-ph/0308036].
%



\bibitem{Contino:2003ve}
  R.~Contino, Y.~Nomura and A.~Pomarol,
  Nucl.\ Phys.\ B {\bf 671}, 148 (2003)
  [arXiv:hep-ph/0306259].
%



\bibitem{Carena:2006bn} See, for example,
  M.~Carena, E.~Ponton, J.~Santiago and C.~E.~M.~Wagner,
  Nucl.\ Phys.\ B {\bf 759}, 202 (2006)
  [arXiv:hep-ph/0607106] and
%
  Phys.\ Rev.\  D {\bf 76}, 035006 (2007)
  [arXiv:hep-ph/0701055]



\bibitem{Contino:2006nn}
  R.~Contino, T.~Kramer, M.~Son and R.~Sundrum,
  JHEP {\bf 0705}, 074 (2007)
  [arXiv:hep-ph/0612180].

\bibitem{Medina:2007hz}
  A.~D.~Medina, N.~R.~Shah and C.~E.~M.~Wagner,
Phys.\ Rev.\  D {\bf 76}, 095010 (2007)
  [arXiv:0706.1281 [hep-ph]].



\bibitem{Serone:2009kf}
  M.~Serone,
  arXiv:0909.5619 [hep-ph].


\bibitem{Hosotani:1983xw}
  Y.~Hosotani,
  Phys.\ Lett.\  B {\bf 126}, 309 (1983) and
  Phys.\ Lett.\  B {\bf 129}, 193 (1983).





\bibitem{Agashe:2004rs}
  K.~Agashe, R.~Contino and A.~Pomarol,
  Nucl.\ Phys.\  B {\bf 719}, 165 (2005)
  [arXiv:hep-ph/0412089];







\bibitem{Hosotani:2008tx}
  Y.~Hosotani, K.~Oda, T.~Ohnuma and Y.~Sakamura,
  Phys.\ Rev.\  D {\bf 78}, 096002 (2008)
  [Erratum-ibid.\  D {\bf 79}, 079902 (2009)]
  [arXiv:0806.0480 [hep-ph]].






\bibitem{Maldacena:1997re}
  J.~M.~Maldacena,
  Adv.\ Theor.\ Math.\ Phys.\  {\bf 2}, 231 (1998)
  [Int.\ J.\ Theor.\ Phys.\  {\bf 38}, 1113 (1999)]
  [arXiv:hep-th/9711200];
%
  S.~S.~Gubser, I.~R.~Klebanov and A.~M.~Polyakov,
  Phys.\ Lett.\ B {\bf 428}, 105 (1998)
  [arXiv:hep-th/9802109];
%
  E.~Witten,
  Adv.\ Theor.\ Math.\ Phys.\  {\bf 2}, 253 (1998)
  [arXiv:hep-th/9802150].




\bibitem{GK}
H.~Georgi and D.~B.~Kaplan,
Phys.\ Lett.\ B {\bf 136}, 183 (1984);
B {\bf 145}, 216 (1984); \\
D.~B.~Kaplan, H.~Georgi and S.~Dimopoulos,
  Phys.\ Lett.\  B {\bf 136}, 187 (1984); \\
H.~Georgi, D.~B.~Kaplan and P.~Galison,
Phys.\ Lett.\ B {\bf 143}, 152 (1984); \\
M.~J.~Dugan, H.~Georgi and D.~B.~Kaplan,
Nucl.\ Phys.\ B {\bf 254}, 299 (1985).



\bibitem{Arkani-Hamed:2000ds}
  N.~Arkani-Hamed, M.~Porrati and L.~Randall,
  JHEP {\bf 0108}, 017 (2001)
  [arXiv:hep-th/0012148];
%
  R.~Rattazzi and A.~Zaffaroni,
  JHEP {\bf 0104}, 021 (2001)
  [arXiv:hep-th/0012248].


\bibitem{Contino:2006qr}
  R.~Contino, L.~Da Rold and A.~Pomarol,
  Phys.\ Rev.\  D {\bf 75}, 055014 (2007)
  [arXiv:hep-ph/0612048].


\bibitem{Carena:2007tn}
  M.~Carena, A.~D.~Medina, B.~Panes, N.~R.~Shah and C.~E.~M.~Wagner,
  Phys.\ Rev.\  D {\bf 77}, 076003 (2008)
  [arXiv:0712.0095 [hep-ph]].
  R.~Contino and G.~Servant,
  JHEP {\bf 0806}, 026 (2008)
  [arXiv:0801.1679 [hep-ph]];
  J.~Mrazek and A.~Wulzer,
  arXiv:0909.3977 [hep-ph];
  %
 for a general analysis of LHC signals of such top quark ``partners'', see
  J.~A.~Aguilar-Saavedra,
  JHEP {\bf 0911}, 030 (2009)
  [arXiv:0907.3155 [hep-ph]].



\bibitem{Chizhov:2009fc}
  M.~V.~Chizhov and G.~Dvali,
  arXiv:0908.0924 [hep-ph].






\bibitem{Agashe:2006at}
  K.~Agashe, R.~Contino, L.~Da Rold and A.~Pomarol,
  Phys.\ Lett.\  B {\bf 641} (2006) 62
  [arXiv:hep-ph/0605341].


\bibitem{Huber:2003tu}
S.~J.~Huber,
Nucl.\ Phys.\ B {\bf 666}, 269 (2003)
[arXiv:hep-ph/0303183];
%
K.~Agashe, G.~Perez and A.~Soni,
  Phys.\ Rev.\ D {\bf 71}, 016002 (2005)
  [arXiv:hep-ph/0408134];
%
  S.~Casagrande, F.~Goertz, U.~Haisch, M.~Neubert and T.~Pfoh,
  JHEP {\bf 0810}, 094 (2008)
[arXiv:0807.4937 [hep-ph]];
%
M.~Blanke, A.~J.~Buras, B.~Duling, S.~Gori and A.~Weiler,
  JHEP {\bf 0903}, 001 (2009)
  [arXiv:0809.1073 [hep-ph]];
%
 M.~Blanke, A.~J.~Buras, B.~Duling, K.~Gemmler and S.~Gori,
  JHEP {\bf 0903}, 108 (2009)
  [arXiv:0812.3803 [hep-ph]];
%
M.~E.~Albrecht, M.~Blanke, A.~J.~Buras, B.~Duling and K.~Gemmler,
  arXiv:0903.2415 [hep-ph];
%
  A.~J.~Buras, B.~Duling and S.~Gori,
  JHEP {\bf 0909}, 076 (2009)
  [arXiv:0905.2318 [hep-ph]].

\bibitem{Csaki:2008zd}
C.~Csaki, A.~Falkowski and A.~Weiler,
JHEP {\bf 0809}, 008 (2008)
[arXiv:0804.1954 [hep-ph]].
%


\bibitem{Agashe:2009tu}
  K.~Agashe,
  arXiv:0902.2400 [hep-ph].



\bibitem{Gedalia:2009ws}
  O.~Gedalia, G.~Isidori and G.~Perez,
  arXiv:0905.3264 [hep-ph].


\bibitem{Fitzpatrick:2007sa}
  A.~L.~Fitzpatrick, G.~Perez and L.~Randall,
arXiv:0710.1869 [hep-ph];
%
  M.~C.~Chen and H.~B.~Yu,
  Phys.\ Lett.\  B {\bf 672}, 253 (2009)
  [arXiv:0804.2503 [hep-ph]];
%
  G.~Perez and L.~Randall,
  JHEP {\bf 0901}, 077 (2009)
  [arXiv:0805.4652 [hep-ph]];
%
  C.~Csaki, C.~Delaunay, C.~Grojean and Y.~Grossman,
  JHEP {\bf 0810}, 055 (2008)
  [arXiv:0806.0356 [hep-ph]];
%
  J.~Santiago,
JHEP {\bf 0812}, 046 (2008)
[arXiv:0806.1230 [hep-ph]];
  C.~Csaki, A.~Falkowski and A.~Weiler,
  arXiv:0806.3757 [hep-ph];
  C.~Csaki, G.~Perez, Z.~Surujon and A.~Weiler,
  arXiv:0907.0474 [hep-ph];
%
  M.~C.~Chen, K.~T.~Mahanthappa and F.~Yu,
  arXiv:0909.5472 [hep-ph].






\bibitem{Davoudiasl:2002ua}
  H.~Davoudiasl, J.~L.~Hewett and T.~G.~Rizzo,
  Phys.\ Rev.\  D {\bf 68}, 045002 (2003)
  [arXiv:hep-ph/0212279];
%
  M.~Carena, E.~Ponton, T.~M.~P.~Tait and C.~E.~M.~Wagner,
  Phys.\ Rev.\  D {\bf 67}, 096006 (2003)
  [arXiv:hep-ph/0212307];
%
  M.~S.~Carena, A.~Delgado, E.~Ponton, T.~M.~P.~Tait and C.~E.~M.~Wagner,
  Phys.\ Rev.\  D {\bf 68}, 035010 (2003)
  [arXiv:hep-ph/0305188];
%
  M.~S.~Carena, A.~Delgado, E.~Ponton, T.~M.~P.~Tait and C.~E.~M.~Wagner,
  Phys.\ Rev.\  D {\bf 71}, 015010 (2005)
  [arXiv:hep-ph/0410344].



\bibitem{soft}
  P.~McGuirk, G.~Shiu and K.~M.~Zurek,
  JHEP {\bf 0803}, 012 (2008)
  [arXiv:0712.2264 [hep-ph]];
%
  G.~Shiu, B.~Underwood, K.~M.~Zurek and D.~G.~E.~Walker,
  Phys.\ Rev.\ Lett.\  {\bf 100}, 031601 (2008)
  [arXiv:0705.4097 [hep-ph]];
%
  A.~Falkowski and M.~Perez-Victoria,
  arXiv:0806.1737 [hep-ph];
%
  B.~Batell, T.~Gherghetta and D.~Sword,
  arXiv:0808.3977 [hep-ph];
  A.~Delgado and D.~Diego,
  Phys.\ Rev.\  D {\bf 80}, 024030 (2009)
  [arXiv:0905.1095 [hep-ph]];
  S.~Mert Aybat and J.~Santiago,
  arXiv:0905.3032 [hep-ph].


\bibitem{private} private communication with
A.~Falkowski and M.~Son.



\bibitem{Group:2009qk}
  T.~E.~W.~Group  [CDF Collaboration and D0 Collaboration],
  arXiv:0908.2171 [hep-ex].




\bibitem{Han:2005ru}
  T.~Han, H.~E.~Logan and L.~T.~Wang,
  JHEP {\bf 0601}, 099 (2006)
  [arXiv:hep-ph/0506313].


\bibitem{Agashe:2007ki}
  K.~Agashe {\it et al.},
  Phys.\ Rev.\  D {\bf 76}, 115015 (2007)
  [arXiv:0709.0007 [hep-ph]];
  K.~Agashe, S.~Gopalakrishna, T.~Han, G.~Y.~Huang and A.~Soni,
  arXiv:0810.1497 [hep-ph].


\bibitem{Falkowski:2006vi}
  A.~Falkowski,
  Phys.\ Rev.\  D {\bf 75}, 025017 (2007)
  [arXiv:hep-ph/0610336].



\bibitem{Agashe:2008uz}
  K.~Agashe, A.~Azatov and L.~Zhu,
  Phys.\ Rev.\  D {\bf 79}, 056006 (2009)
  [arXiv:0810.1016 [hep-ph]].


\bibitem{Pumplin:2002vw}
  J.~Pumplin, D.~R.~Stump, J.~Huston, H.~L.~Lai, P.~M.~Nadolsky and W.~K.~Tung,
  JHEP {\bf 0207}, 012 (2002)
  [arXiv:hep-ph/0201195].

  \bibitem{cuts}
  G.~L.~Bayatian {\it et al.}  [CMS Collaboration],
  J.\ Phys.\ G {\bf 34}, 995 (2007).
  G.~Aad {\it et al.}  [The ATLAS Collaboration],
  arXiv:0901.0512 [hep-ex].

\bibitem{Abazov:2007ev}
  V.~M.~Abazov {\it et al.}  [D0 Collaboration],
  Phys.\ Rev.\ Lett.\  {\bf 99}, 191802 (2007)
  [arXiv:hep-ex/0702005].










\bibitem{Han:2003wu}
  T.~Han, H.~E.~Logan, B.~McElrath and L.~T.~Wang,
  Phys.\ Rev.\  D {\bf 67}, 095004 (2003)
  [arXiv:hep-ph/0301040].




\bibitem{Keung:1983uu}
  W.~Y.~Keung and G.~Senjanovic,
  Phys.\ Rev.\ Lett.\  {\bf 50}, 1427 (1983).







\end{thebibliography}
\end{document}